\RequirePackage{etoolbox}

\newtoggle{paper}
\newtoggle{report}

\toggletrue{paper}

\documentclass[sigplan,screen,authorversion]{acmart}

\usepackage{listings}
\usepackage{multicol}
\usepackage{subcaption}
\usepackage{wrapfig}
\usepackage{enumitem}
\usepackage{multirow}
\usepackage{adjustbox}
\usepackage{mdwlist}
\usepackage{soul}
\usepackage{xcolor}
\usepackage{stackengine}
\setlist[description]{leftmargin=2\parindent,labelindent=\parindent}

\lstdefinelanguage{plain}
{
  sensitive=true
}
\lstdefinestyle{diff}
{
  moredelim=*[il][\color{ACMGreen}]{++},
  moredelim=*[il][\color{ACMRed}]{--},
}

\newcommand{\code}[1]{\mbox{{\ttfamily #1}}}

\acmDOI{10.1145/3640537.3641578}
\acmYear{2024}
\copyrightyear{2024}
\acmSubmissionID{cc24main-p91-p}
\acmISBN{979-8-4007-0507-6/24/03}
\acmConference[CC '24]{Proceedings of the 33rd ACM SIGPLAN International Conference on Compiler Construction}{March 2--3, 2024}{Edinburgh, United Kingdom}
\acmBooktitle{Proceedings of the 33rd ACM SIGPLAN International Conference on Compiler Construction (CC '24), March 2--3, 2024, Edinburgh, United Kingdom}
\received{2023-11-13}
\received[accepted]{2023-12-23}

\setcopyright{acmlicensed}

\newcommand{\mynote}[2]{
  \fbox{\bfseries\sffamily\scriptsize#1}
    {\color{red}\small\textsf{\emph{#2}}}
}

\ifx\comments\undefined
\newcommand{\todo}[1]{}
\newcommand{\fix}[1]{}
\newcommand{\sk}[1]{}
\newcommand{\jrs}[1]{}
\else
\newcommand{\todo}[1]{\mynote{TODO}{#1}}
\newcommand{\fix}[1]{\mynote{FIX}{#1}}
\newcommand{\sk}[1]{\mynote{SK}{#1}}
\newcommand{\jrs}[1]{\mynote{JRS}{#1}}
\fi

\newcommand{\ifonecolumn}[2]{\ifnum\columnwidth=\textwidth#1\else#2\fi}

\newcommand{\mi}[1]{\mathit{#1}}

\def\DWARF{\textsc{Dwarf}}

\makeatletter
\renewcommand{\sectionautorefname}{\S\@gobble}
\renewcommand{\subsectionautorefname}{\S\@gobble}
\renewcommand{\subsubsectionautorefname}{\S\@gobble}
\makeatother

\begin{document}

\title{Accurate Coverage Metrics for Compiler-Generated Debugging Information}

\author{J. Ryan Stinnett}
\orcid{0000-0002-3101-1189}
\affiliation{
  \institution{King's College London}
  \city{London}
  \country{United Kingdom}
}
\email{jryans@gmail.com}

\author{Stephen Kell}
\orcid{0000-0002-2702-5983}
\affiliation{
  \institution{King's College London}
  \city{London}
  \country{United Kingdom}
}
\email{stephen.kell@kcl.ac.uk}

% All submissions must be made electronically through the conference submission
% website and include an abstract (100–400 words), author contact information,
% the full list of authors and their affiliations. Full paper submissions must
% be in PDF formatted printable on both A4 and US letter size paper.

% All papers must be prepared in ACM Conference Format using the 2-column acmart
% format: use the options \documentclass[sigplan,10pt,review,anonymous]{acmart}
% for LaTeX, and interim-layout.docx for Word. Important note: The Word template
% (interim-layout.docx) on the ACM website uses 9pt font; you need to increase
% it to 10pt.

% Papers should contain a maximum of 10 pages of text (in a typeface no smaller
% than 10 point) or figures, NOT INCLUDING references. There is no page limit
% for references and they must include the name of all authors (do not use et
% al.).

% Appendices are not allowed, but the authors may submit anonymous supplementary
% material, such as proofs, source code, or data sets; all supplementary
% material must be in PDF or ZIP format. Looking at supplementary material is at
% the discretion of the reviewers.

\begin{abstract}
  Many debugging tools rely on compiler-produced metadata to present a
source-language view of program states, such as variable values and source line
numbers.
While this tends to work for unoptimised programs,
current compilers often generate only partial debugging information
in optimised programs.
Current approaches for measuring the extent of coverage of
local variables are based on crude assumptions
(for example, assuming variables could cover their whole parent
scope) and are not comparable from one compilation to another.
In this work,
we propose some new metrics, computable by our tools, which could serve as
motivation for language implementations to improve debugging quality.

\end{abstract}

%% 2012 ACM Computing Classification System (CSS) concepts
%% Generate at 'https://dl.acm.org/ccs'.


\maketitle

\section{Introduction}

Compilers emit debugging metadata, or ``debug info'', to
enable the mapping of machine program states
back to source program states.
This information is consumed not only by interactive debuggers, but also by
profilers, tracers (e.g.\ SystemTap \shortcite{eiglerSystemTap2006}), coverage tools, etc.
Toolchains have developed standard formats for it,
such as \DWARF{} \shortcite{dwarfDebuggingInformation2017},
which decouple tools from compilers.
Unfortunately,
\emph{compiler optimisations} interact poorly with debug info:
the information produced by production-grade compilers is often incorrect or missing
after optimisation
\cite{
  hennessySymbolicDebuggingOptimized1982,
  zellwegerInteractiveHighlevelDebugger1983a,
  coutantDOCPracticalApproach1988,
  brooksNewApproachDebugging1992,
  ticeOPTVIEWNewApproach1998,
  jelinekImprovingDebugInfo2010,
  olivaGCCGOlogyStudying2019,
  liDebugInformationValidation2020,
  dilunaWhoDebuggingDebuggers2021,
  assaianteWhereDidMy2023}.

When this happens, the tool is unable to identify correctly
the source-level program state of interest.
In debuggers, a common failure of this kind
is the message ``variable optimised out''
when attempting to print or evaluate a local variable.
This message often occurs even when a variable remains represented in the program;
it is triggered when the debug info, not the variable, is missing.
\sk{Anecdata about (1) extent of debug info elimination outstripping, and (2) bug reports being fixed}
When coverage is lacking like this, the effectiveness of the tooling is degraded.

Debugging information for optimised code matters.
Some codebases cannot be built without optimisation (e.g.\ the Linux kernel),
and others cannot meaningfully be run without it (e.g.\ resource-heavy programs
such as games which are unusable without optimisation).
Some bugs occur only on optimised ``full speed'' code,
and some tools are only useful on the same
(e.g.\ profilers).
Programmers are used to deploying partial workarounds when facing difficulties
debugging optimised code---notably,
rebuilding without optimisation---but this
%diminishing its usefulness and trustworthiness.
brings costs to developers (e.g.\ rebuilding takes time)
and does not help in-the-field bug reporting by end users
(being optimised, deployed binaries are frequently undebuggable).

Whereas compiler benchmarks provide a basis for evaluating
the optimisation benefit achieved for a program,
there is no equivalent way to measure (or show absence of) the incurred \emph{debuggability disbenefit}
in the generated ``program plus debug info''.
Certain crude metrics do exist but, as we will survey,
they have various flaws and, most notably,
are not comparable across compilers.
One reason for this is that
although missing debug info is usually regarded as a compiler bug,
it is currently not clear what it means for debug info to be fully
complete and correct.
Historically, a ``best effort'' approach has prevailed,
pursuing specific improvements \cite{brenderDebuggingOptimizedCode1998} but
placing no strong correctness criterion on compilers.
%Compiler bugs are sometimes deemed ``fixed'' simply by suppressing incorrect debug info,
%rather than by ensuring that info is both correct and complete\sk{want to forward-reference
%some anecdotes here}.

% recent work by \citet{assaianteWhereDidMy2023} has observed the difficulty distinguishing
% ``unavoidable effect[s] of optimization'' from fixable bugs,
% and remarked that debug info completeness has ``no reliable oracle''.
% \sk{Still a bit WIP, but I feel it might work to elaborate this into ``weak''
% and ``strong'' correctness criteria, where the ``strong'' requires \DWARF{}
% extensions for residualisation, but ``weak'' does not.}

In this paper we develop
the first robust approach to measuring the coverage of \emph{local variable information}
in \DWARF{} debugging information.
Our contributions are the following:
\begin{itemize}

  \item a model of optimisation under debugging as ``residualising''
  computation, with an analysis of local variables
  and their life-cycle within a debuggable program (\autoref{sec:life-cycle});

  \item a discussion of a series of candidate \emph{local variable coverage}
  metrics (\autoref{sec:illusion-coverage-defined}),
  with experience of applying them to complex cases
  found in real debug info,
  culminating in an implemented coverage tool which improves on earlier metrics
  by obtaining an accurate and achievable ``complete coverage'' baseline
  (\autoref{sec:implementation});

  \item experimental evidence showing that our metric
  reliably reflects changes in debuggability across compiler versions
  (\autoref{sec:metrics}),
  can explain the debuggability effects of
  compiler changes (\autoref{sec:case-studies})
  and can reproduce \emph{with expected differences}
  findings of a prior study (\autoref{sec:replication}).
\end{itemize}

Our experiments and implemented tools are available
in a deposited artifact \cite{debugInfoMetricsArtifact} (not peer-reviewed).

\section{Understanding Debug Coverage}
\label{sec:variable-coverage}
What kinds of coverage could we measure, and what distinguishes coverage from correctness?
We discuss these here.

\subsection{Distinguishing Coverage from Correctness}

Given ``full'' debug info known to be complete and correct,
we would expect it to satisfy the following properties:

\begin{description}

\item[Control coverage] It is possible to stop at all
reachable points in the source program's control flow. Put differently,
if code actually executes, then it also appears to execute
as observed from the debugger.\footnote{From here on we assume
an imperative source language, i.e.\ one with explicit control flow,
although we believe this could be generalised---perhaps in terms of
a partial order on active program constructs.}

% Maybe clarify : "stepping on each line yields the complete set of reachable source lines,
% albeit possibly in a surprising order and with duplicates"

\item[Variable coverage] At any such point, it is possible to examine
all named values that the source program deems to be \emph{in scope} at that point
and also taking a \textsf{well-defined} value---roughly, it is not uninitialized.
(From here we use ``variable'' to refer to named values
even if the value happens to be immutable,
such as a \code{const} local in C.)

% is "truthful" just "simulation up to input/output behaviour" and
% "expected" just "simulation up to program order"?

\item[Semantic consistency] When examining a variable at such a point,
its observed value should be (somehow) consistent with the source program's
semantics.

\end{description}

Whereas the first two properties (coverage) are reasonably precise,
semantic consistency can be seen as a spectrum of possible correctness conditions:
%Semantic consistency properties can be seen as
%\emph{correctness} properties on debug info, as distinct from coverage properties.
%Clearly they form a spectrum:
exactly \emph{how consistent} one should expect observations to be
might be answered in strong terms (``as if executing the source program unoptimised'')
or relatively weaker terms, e.g.\ permitting optimisers to reorder code.
(An analogy exists here with memory models,
whose consistency properties may be stronger or weaker.
Prior work \citep{zellwegerInteractiveSourcelevelDebugging1984} has given names to
two polar-opposite properties:
``expected'' means that even when optimisations are
enabled, the debug-time view appears identical to unoptimised execution,
while ``truthful'' means that the debugger shows whatever states are actually
inhabited by the optimised program, translating them
into source terms as best it can but applying no particular correctness criterion.)
% To our knowledge, the ``expected'' property
% is not offered by any optimising ahead-of-time
% compiler toolchain using debug info formats such as \DWARF{} \shortcite{dwarfDebuggingInformation2017}.)
% Rather, it appears only in interpreters---either
% simple interpreters, or in sophisticated language virtual machines' debug servers
% that use dynamic deoptimisation \citep{holzleDebuggingOptimizedCode1992}
% to switch to an interpreter\footnote{This extends to
% template-based just-in-time compilers, which, like interpreters,
% also preserve a direct simulation of the source program.} on demand.
% In both cases, the debugging infrastructure never observes optimised code,
% unlike toolchains using \DWARF{} or the like.
% We focus on the toolchain scenario in this work.

In this work we focus on measuring the \emph{extent} of local variable debug info (coverage)
but not its semantic consistency (correctness).
%meaning the choice of a particular semantic consistency property is also out of scope.
Although this means a compiler could game our metric by adding spurious debug info,
developers are unlikely to add this intentionally since it
would offer an especially bad user experience for end developers.

\ifonecolumn{
\begin{wrapfigure}[12]{r}{0.45\linewidth}
\vspace*{-3ex}
}{
\begin{figure}[thb]
}

{\scriptsize\begin{lstlisting}[language=plain,basicstyle=\ttfamily,columns=flexible]
0xb90..0xbb3: (reg RDX)
0xbb3..0xbdb: (value (div (- (reg RAX) (reg RDI)) 4))
0xbdb..0xc0e: (reg RDX)
0xc0e..0xc1a: (frame_offset -24)
\end{lstlisting}}
\ifonecolumn{\vspace*{-1ex}}{}
  \caption{How \DWARF{} might describe a local variable
 over four distinct address ranges within a function.
   Most expressions compute where it is \emph{located}: in a register or (later) on the stack.
  Over the second range, however, it is not represented explicitly; its \emph{value} is computed
  as a scaled difference of two registers.
  (The textual syntax is for illustration only.)}
  \label{fig:dwarf-exprs-example}
\ifonecolumn{
\end{wrapfigure}
}{
\end{figure}
}
% this example is based on the 'envp' argument (DIE 0x1d83) from the '_init' subprogram
% in srk's build of glibc 2.31-13+deb11u5 (build ID b5/03275bf9fee51581fdceef97533b194035b4f7)

\subsection{Debug Info as Residualised Code}
\label{sec:value-expressions}

Compilers may optimise code so that a variable
is no longer explicitly represented.
Such variables remain coverable at debug time
in modern debug info formats like \DWARF{} \shortcite{dwarfDebuggingInformation2017},
which describe them as functions to be computed by the debugger.
In \DWARF{} these are \emph{expressions} in an interpreted stack machine language.
Expressions can compute a variable's \emph{location} in memory or the register file---perhaps
a simple offset from the stack pointer,
but sometimes complex (e.g.\ in a nested function, traversing a static-link pointer to reach
locals in the lexically enclosing scope)---or
they can compute a variable's \emph{value} directly.
\autoref{fig:dwarf-exprs-example} shows an example.

Compilers may also optimise code so that an intermediate
control-flow position of the source program is elided,
having no corresponding program counter position in the object code.
Again, full variable coverage over these positions often remains feasible,
as debug info can represent intermediate states to be synthesised by the debugger,
%e.g.\ during linewise stepping.
%These#
effectively existing \emph{in between} the instructions of the object program.
\autoref{fig:location-views-example} shows an example.
This facility is less well established
but is implemented in (at least) extensions to \DWARF{}
that are proposed for the next standard and already used by GCC
\cite{olivaConsistentViewsRecommended2010,
olivaStatementFrontierNotes2017,
olivaLocationViewNumbering2017}.
Currently, eliminated \emph{branching} control flow cannot be represented
in \DWARF{} or any proposed extension.
% we return to this later in \autoref{sec:eliminated-branching-control-flow}.)

\ifonecolumn{
\begin{wrapfigure}[12]{r}{0.45\linewidth}
\vspace*{-3ex}
}{
\begin{figure}[bht]
\vspace*{-1ex}
}

% {\tiny\begin{lstlisting}[language=plain,basicstyle=\ttfamily,columns=fixed]
%    source     | instructions         | location views, value exprs
% ------------------------------------------------------------------
% 1: int x = 1; |                     / view 1: ln 1, x: (value 1)
% 2: x++;       |                    /  view 2: ln 2, x: (value 1)
% 3: x++;       |                   /   view 3: ln 3, x: (value 2)
%               | 0xd06: mov $3,%rdi___ view 4: ln 4, x: (value 3)
% 4: x = f(x);  | 0xd0d: callq f                ln 4, x: (value 3)
%               | 0xd12: mov %rax,%rdi |        ln 4, x: (reg RAX)
% 5: ...        | 0xd15: ...           |        ln 5, x: (reg RDI)
% \end{lstlisting}}
\begin{center}
% Use https://app.diagrams.net to edit the SVG, then export to HTML, view in
% browser, and save as PDF
\includegraphics{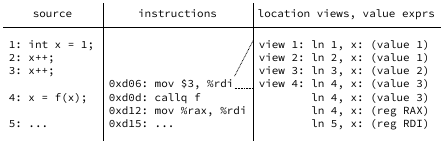}
\end{center}
\ifonecolumn{\vspace*{-1ex}}{}
  \caption{How \DWARF{} can conceptually residualise control-flow positions
  that were eliminated during optimisation,
  using the ``location views'' extension of
  \citet{olivaConsistentViewsRecommended2010}.
  A single program counter value (here \texttt{0xd06}) can have a numbered
  sequence of ``views'' such that a single local variable is described differently for each view.
  At debug time, control appears to pass through each view in sequence, even though
  the program counter does not advance.\sk{Diagram is placeholder}}
  \label{fig:location-views-example}
\ifonecolumn{
\end{wrapfigure}
}{
\end{figure}
}

\label{sec:residualisation-concept}
Both of these can be thought of as a kind of ``residual computation'':
code eliminated from the program is in effect reinstated in the debug info.
We view compilers as outputting two artifacts:
the object program proper, and its debug info.
Elimination from one does not imply elimination from the other,
%In fact the two are in opposition:
%in reaction to how the latter was optimised.
and in fact the opposite is true:
the more thoroughly a variable is ``eliminated'' from the emitted program,
the more it needs to be \emph{described} in the debug info.
Any notion of full coverage needs to reflect the potential for residualisation:
it is a loss of coverage if the compiler did not take the opportunity to residualise
a local variable over some reachable range of positions in the source program.
Given adequate residualisation features,
optimisations and debugging need not be mutually excluding.

%we see debug info as \emph{residualising} the eliminations or simplifications
%made during optimisation, i.e.\ describing
%how to undo the effect of optimisations when observing the optimised program.
%We return to this idea throughout the \work{}.
%In a toolchain that supports ``truthful'' debugging,
%optimisation is best seen conceptually as residualisation,
%moving code (or even state) out of the program proper and into its debug info,
%rather than outright simplification or elimination.

\ifonecolumn{
\begin{wrapfigure}[16]{r}{0.35\linewidth}
}{
\begin{figure}[b]
}
  \ifonecolumn{\vspace*{-2ex}}{}
  \includegraphics[width=0.28\textwidth]{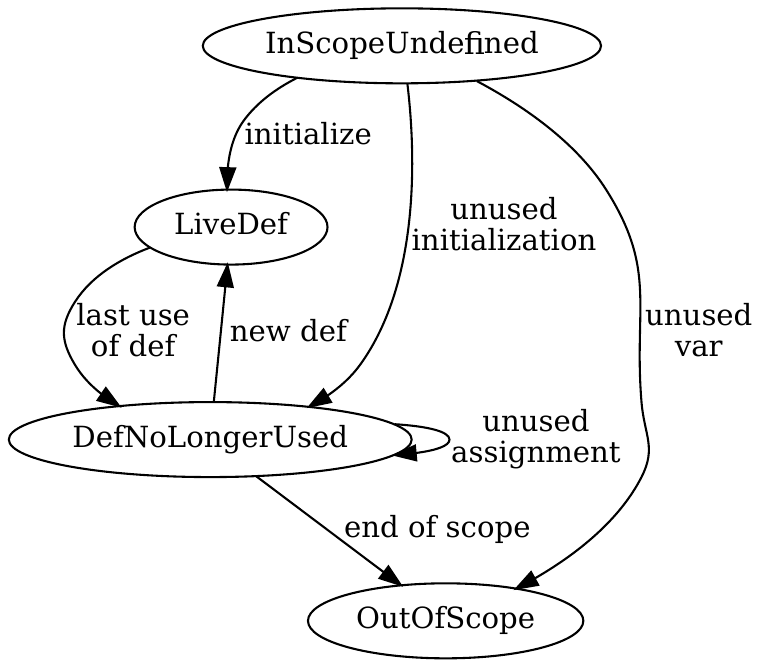}
  \caption{Classical view of the life cycle of a local variable.}
  \label{fig:def-use-life-cycle}
\ifonecolumn{
\end{wrapfigure}
}{
\end{figure}
}

\subsection{Research Questions}
\label{sec:research-questions}

Using this perspective of debug info \emph{residualising}
otherwise eliminated or simplified computation,
we can now state the specific research questions
to be answered in this paper.

\textbf{RQ1. For measuring coverage, what is a viable conceptual basis?}
Clearly coverage can be measured as a fraction of what is \emph{covered}
relative to what is \emph{coverable}, but how to determine these is unclear.
More specifically: given a program and its debug info,
what configurations may a variable $v$ inhabit at a program point $p$?
%and how do these determine whether it is \emph{covered} and \emph{coverable} at that point?
We will develop a case analysis to answer this question
(\autoref{sec:case-analysis}).

\textbf{RQ2. For measuring coverage, what are the necessary practical steps?}
How can we measure the extent of local variable coverage
in debug info, in a way that fairly represents the effective debug-time availability
of the program's local variables?
To answer this, we need to define an accurate baseline for
when a variable \emph{should} be present, i.e.\ one that
accounts for residualisation opportunities.
This is complicated by the sometimes
unpredictable and counterintuitive behaviour
of compilers and the partial information they generate.
%  such as
% %common subexpression elimination (which can lead to one piece of machine code performing ``dual duty''),
% strength reduction (after which a local variable might no longer be directly represented),
% inlining (which mixes otherwise unrelated source scopes
% into the same chunk of object code)
% and so on.
We define a measure of coverage with respect to a given local variable,
capturing over how much of the program the debugger is able to show a value
for it, relative to the achievable baseline.
% as a fraction of the program locations for which the variable is \emph{in-scope and defined}.
We improve on previous metrics by ensuring
the baseline \emph{is} achievable,
using both conceptual insights (\autoref{sec:illusion-coverage-defined})
and experiences during implementation (\autoref{sec:implementation}).

\textbf{RQ3. In aggregate, how does our metric's picture of debuggability
depart from those computed by na\"ive variations and/or previously proposed metrics?}
We will answer this open question
both by applying our metric to both real and synthetic programs
and exploring the results in plots,
and by a \emph{replication study}
in which we use our metrics to reproduce
an experiment from recent literature.

\textbf{RQ4. In detail, does our metric explain debuggability gaps
in a way consistent with how these are understood by real compiler developers?}
We expect low scores in our metric to be indicative of compiler
bugs, and higher scores to be indicative of their absence.
We study two real fixed compiler bugs
% (selected from a larger survey of
% 25 existing LLVM issues and 6 LLVM community discussions)
before and after their fixes
and explore how these are reflected by our metric\sk{ideally make this more precise}.

\section{A Conceptual Basis for Coverage Metrics}
\label{sec:life-cycle}
\label{sec:case-analysis}

% To define coverage robustly,
% %understand what it means for local variables to be \emph{covered} by debug info,
% we need a conceptual model of source-level variables
% as embodied partly as an object program and partly residualised into debug info.

At a high level, measuring coverage means computing the following:
\[
\mi{coverage} =  \frac{\#\mi{covered}}{\#\mi{coverable}}
\]
This demands answers to three questions:
what counts as ``covered'',
what counts as ``coverable'',
and in what unit these are counted.
In this section we answer the first two of these,
by proposing definitions
for when a local variable is \emph{covered} and \emph{coverable},
made by analogy with a familiar liveness analysis using the data-flow method.

% a program and its debug info, what configurations may a
% variable v inhabit at a program point p, and how do these
% determine whether it is covered and coverable at that point?
% We will develop a case analysis to answer this question.

\subsection{Liveness as a State Space}
Traditional liveness analysis
deems a variable to be live at a point if a \emph{definition} reaches that point
and has at least one later \emph{use}.
We can model each variable in such a scenario
by two unary predicates:
whether it has been initialized (``Defined'' or $D$)
and whether the current definition will be used again (``Live'' or $L$).
Although these two predicates are orthogonal,
a reduction applies since conventionally ``Live'' is assumed to imply ``Defined''
(i.e.\ uninitialized reads are not considered), leaving three in-scope states.
The resulting machine is shown in \autoref{fig:def-use-life-cycle};
for readability, an additional ``out of scope'' exit node is added.
%(For readability,
%but this falls outside the 2-bit state space described.
%A third \emph{in-scope} bit could be added but would be largely redundant.)

%\subsection{Modelling debug coverage of a variable}
\subsection{A More Realistic Life Cycle}
\label{sec:debug-info-coverage}

% "short name" used to say: in Fig.\ \ref{fig:full-life-cycle}
\begin{table*}[htb]
\caption{Case analysis of in-scope variables, which at any point may ($A$) be allocated
  to a storage location, ($D$) take a defined value according to source semantics,
  ($L$) hold a program-live value, and
  ($K$) be possibly knowable given the right debug info.}
\begin{tabular}{|l|l|l|l|p{0.16\linewidth}|p{0.62\linewidth}|}\hline
 $A$& $D$ &$K$ &$L$ &   short name  & example/notes\\\hline
  0 &  0  & 0  & 0  &  InScopeOnly   & just come into scope; neither allocated nor defined\\
  0 &  1  & 0  & 0  &  Unknowable    & no storage, value not program-live, no longer recoverable from other state\\ %<--- knowledge extension
  0 &  1  & 1  & 0  &  KnowablePDead & not program-live nor allocated, but recoverable as a function of other state \\ %<-----common BUG: compilers don't emit these
  0 &  1  & 1  & 1  &  UnallocatedPLive & normal case of non-allocated live variable\\ %<-----also common BUG: compilers don't emit these
  1 &  0  & 0  & 0  &  AllocatedUninit & uninitialized reg or stack slot\\
  1 &  1  & 0  & 0  &  AllocatedStale  & program-dead store eliminated; storage still allocated\\ %<---- BUG!
  1 &  1  & 1  & 0  &  AllocatedPDead &  program-dead variable but correct value still stored\\ %<-----common BUG: compilers don't emit these
  1 &  1  & 1  & 1  &  NormalPLive    & normal case: program-live, allocated variable\\\hline
\end{tabular}
\label{tab:life-cycle}
\end{table*}

In a debugging scenario, the situation is more complex
in a few ways.

\paragraph{Multiple deaths} As before, a variable may be defined (initialized) or not,
but ``dead'' is a less clear concept.
An in-scope variable that is deemed dead by the optimiser
may still be requested by the debugger's user, even though the program does not need it.
From here we qualify ``dead'' as ``program-dead'', to highlight this.
Consider the straight-line program shown in \autoref{fig:liveness-diagram}.
On the right is the source code,
and on the left is a flow chart of significant events in the variable's lifecycle,
showing its two additional deaths: when it is unrecoverable (unknowable),
and when it is no longer in scope.

\paragraph{Allocated vs residual} Unlike a traditional data-flow analysis
we care to distinguish whether optimisation passes have
residualised a source variable into a debug info expression
or simply represented it directly in some allocated storage (a register or stack slot).

\paragraph{Actual versus potential} Variables are routinely residualised into expressions,
but imperfect debug info may miss opportunities to do so.
It matters whether a source variable \emph{can be} expressed in this way or not.
(This links back to ``multiple deaths'': one of the two
``later deaths'' is when there is no such function
of program state.)

%``dead'' needs to be refined to reflect a difference in recoverability.

To account for all these distinctions, we can
identify the following largely orthogonal predicates
of an in-scope variable:

\begin{itemize}
  \item Is it ``allocated'' to any storage location ($A$)? (If not, it might still be computed by an expression.)
  \item Is it initialized i.e.\ ever-defined ($D$), according to source semantics?
  \item Is its current definition [program-]live ($L$)?
  \item Does it have a value that is knowable ($K$) from the current object program state,
  i.e.\ as any expression over it?
\end{itemize}

% \sk{REMOVE ME}
% \begin{figure}
% \includegraphics[width=0.8\textwidth]{figures/full-life-cycle.eps}
% \caption{How variables may transition
% between the states enumerated in \autoref{tab:life-cycle}.
% For readability, transitions are omitted if modelled by spending zero time in an
% intervening state. For example, a variable might immediately
% transition from RecoverablePDead to OutOfScope, but this can be modelled
% as occupying Unknowable over a zero-length intermediate range of program locations.\sk{Unsure whether this
% figure is good value.}}
% \label{fig:full-life-cycle}
% \end{figure}
% a variable may be ``defined in program state'' or perhaps
% ``defined only residually'', i.e.\ in the form of an expression.
%
% This most often happens when the variable is ``dead''
% and the register allocator has recycled its storage.
%
% Optimisation passes may have
% residualised a source variable into a debug info expression,
% so a variable may be ``defined in program state'' or perhaps
% ``defined only residually'', i.e.\ in the form of an expression.
% Or a variable might not be described in debug info at all---hopefully
% because its value is no longer represented in any way,
% but perhaps from a coverage bug where a viable expression does exist but simply was not generated
% by the compiler.

\begin{figure*}
\begin{multicols}{2}
  % Use https://app.diagrams.net to edit the SVG, then export to HTML, view in
  % browser, and save as PDF
  \includegraphics{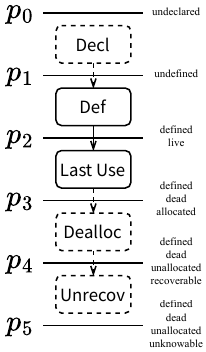}
\columnbreak
  {\footnotesize\begin{lstlisting}[language=C,basicstyle=\sffamily,columns=fullflexible,gobble=4]
    int example(int n) {
      // p0: before `x` enters scope (undeclared)
      int x;
      // p1: before live range of `x` (undefined)
      x = 1;
      // p2: within live range of `x` (defined, live)
      int b = x + 2;
      // p3: after live range of `x` (defined, program-dead, allocated)
      ...
      // p4: later, once `x`'s storage is re-used (defined, p-dead...
      f(b); // ... , unallocated, but `x` is recoverable as b - 2)
      // p5: later still, `x` is no longer recoverable...
      // (defined, program-dead, unallocated, unknowable)
      return 0;
    }
  \end{lstlisting}}
\end{multicols}
  \vspace*{-4ex}
  \caption{Variable $x$ is in different states across a range of
  program points \(p\). The right-hand C source function is annotated with these points.
  The left-hand flow chart illustrates
  $x$'s lifecycle in a hypothetical compilation,
  from before $x$'s introduction (Decl), across its
  definition (Def), last use (Last Use) and the subsequent reclamation of its stack or register
  storage (Dealloc). A final Unrecov event models the point from which
  there is no longer enough state to reconstruct the value.% as a function of other state.
  }
  \label{fig:liveness-diagram}
\end{figure*}

These four elements are again somewhat non-orthogonal.
To be knowable and to be live both require being defined.
(We regard uninitialized values as meaningless and therefore trivially unknowable.)
To be allocated and program-live is also to be knowable
(simply by loading from the allocated storage;
a \DWARF{} ``location expression'' denotes this).
After applying these reductions, eight states of a possible 16 remain;
the state space is shown in \autoref{tab:life-cycle}.
%and \autoref{fig:full-life-cycle}.
\sk{to reinstate full life cycle diagram?}

``Knowable'' here means that reading from storage and/or
evaluating \emph{some} debug info expression \emph{could} obtain the variable's value.
This subsumes three sub-cases:
the allocated case (``knowable'' simply by loading from storage),
the ``recoverable'' case when a variable is
computable as a function of other (redundant) state
(as the ``scaled difference of pointers'' in \autoref{fig:dwarf-exprs-example}),
but also the ``literal value'' case e.g.\ in

\code{int~i~=~0;~...~i = ...;}

\noindent\ldots{} where from the initialization,
the debug info simply records the literal zero,
%Over that range,
no storage needs to be allocated, and
no ``function'' of program state \textit{per se} is required.

The restriction of $K$ to functions of the \emph{current}
object program state reflects how under current tools, when state is to be dropped from the
object program, there is no mechanism that can
cause it to be remembered e.g.\ by an attached debugger.
Dropped state is not residualised.
%\footnote{Of course,
%\emph{omniscient} or time-travel debuggers do not suffer this problem,
%but are forced to take a very different implementation approach
%in order to systematically retain all state.}
Whether there exists any such function appears undecidable in general,
\sk{Why do I think this? Can I reduce it to the halting problem?}
posing a problem for our metric---which seeks to determine
whether each variable is possibly cover\emph{able}.
For now, we assume that all local variables are coverable
at all points where they are defined,
but with optional special handling for the positions
after their last use (i.e.\ when they are finally program-dead).
This can avoid penalising compilers for discarding state as a usual register allocator would,
with the rationale that this \emph{could} still be covered by a debugger
using a technique we describe later (\autoref{sec:knowledge-extension}).
\sk{We actually do an optimistic `covered' rather than a reduction to `coverable'}

Finally we define two further predicates:

\begin{itemize*}
\item $S(p)$ is true iff a variable is in scope at $p$;
\item $B(p)$ is true iff the debugging information descri\ul{\emph{b}}es a variable at $p$.
\end{itemize*}
\noindent{}Note that $B$ is true when the debug info \emph{actually does} describe the variable,
either in its allocation or as a residualised expression---as distinct from
whether it potentially could ($K$).
As in \autoref{fig:def-use-life-cycle}, when a variable is out of scope, the other predicates are no longer of interest.
%\footnote{One could,
%however, imagine debug info that describes variables that happen to remain \emph{knowable} even \emph{after}
%the end of their scope. In this paper we do not seek to measure this kind of ``bonus coverage''.}

From here, for notational convenience
we will identify a predicate with the set of program points where it is true.
We can write $B_v(p)$, say, to refer to whether variable $v$ is described at a point $p$,
or we can write simply $B_v$ to refer to the subset of program points $p \in P$ where $B_v(p)$ is true.
Here $P$ is the set of all program points as somehow defined;
exactly what should constitute a ``program point'' will be discussed in the next section.

% for an optional  variation
%
%  an uncertainty about whether program-dead variables that are \emph{uncovered}
% are in fact \emph{uncoverable}:
% and therefore any such point where they are unknowable under given debug info,
% is to be counted as a loss of coverage.
% \sk{Am thinking we need to defend this earlier and harder.
% By appearing only here it may come as a nasty surprise to the reader.}
%
% This highlights an uncertainty about whether program-dead variables that are \emph{uncovered}
% are in fact \emph{uncoverable}:
% when a value is overwritten in a register or memory,
% can it be recovered as a function of other state?
% If there is no such function, the variable should properly be excluded from what is \emph{coverable}
% and therefore not lessen the coverage metric.
% If, however, there does exist such a function,
% the opposite should be true.
% %it should be deemed coverable,
% %and omitting to cover it should lessen the metric.

\section{Defining a Coverage Metric}
\label{sec:illusion-coverage-defined}

In this section we begin with
metrics already found in the literature,
progressively identifying and eliminating issues
with them,
culminating in our proposed metric.

\subsection{Na\"ive Instruction-Based Metrics}
\label{sec:instruction-based}

A na\"ive but easily computed metric simply counts
the number of instruction bytes over which the value is described ($B$) by debug info,
as a fraction of a total possible number of instruction bytes over which it is in scope ($S$).

\vspace*{-3.0ex}
\[
C_{v} =  \frac{\bigm|B_v\bigm|}
              {\bigm|S_v\bigm|}        \text{~~for~~} P \text{~~the program's set of instruction bytes}
\]
\vspace*{-2.0ex}

\sloppypar{}This is what is computed by existing tools such as \code{llvm-dwarfdump}
\cite{llvmprojectLlvmdwarfdumpDumpVerify2020}
and \code{debuginfo-quality}
\cite{ocallahanComparingQualityDebug2018}.
Unfortunately this results in four problems.

Firstly, measurements are \textbf{not comparable} across compilers,
or even across differently configured runs of the same compiler.
Instruction counts reflect details of the compilation,
such as how many instructions were used to realise a particular feature
of the source program,
which vary independently of how debuggable the result is.
%Numbers of instructions have little to do with
%the source-level debugging experience;
Essentially this metric places a varying, compilation-specific weighting on
the coverage available across parts of the program, according to how many instructions
the compiler used to realise them.

Secondly, in the presence of \emph{location views},
as shown in \autoref{fig:location-views-example},
it simply \textbf{omits to count some coverage}.
Location views can cover a source variable
over a range of zero instructions, which by definition
will not be counted.

Thirdly, in practice it \textbf{accidentally favours some compilations},
specifically unoptimised compilations
which put local variables in a stack slot for their whole lifetime.
\label{sec:register-allocation}
Conversely it penalises optimised compilations using a register allocator.
This is because a variable described as being located in a stack slot will be in $B$ (it is described)
over all program points in $S$, even those where the variable is not yet defined (points not in $D$).
Conversely, a variable that is stored only in registers
will be in $B$ only where it is also in $D$
because registers are allocated not over a whole function
but over specific ranges of instructions---\emph{live ranges},
which always start at a \emph{definition}, so naturally
exclude program points not in $D$.
%Using $B$ as the numerator
%means the stack case will be viewed as `covered' because a location is described,
%even over instructions in which that location contains no meaningful value.
Put differently, it overestimates coverage
by counting as covered a variable in a described stack slot that holds only
garbage (uninitialized) data ($B \supset D$).
(Usually a ``variable not defined'' message would be preferable in these cases!)
This overestimation means \textbf{full coverage becomes impossible
to achieve} when using register allocation,
since the denominator counts program points
over which the variable will have no described value
so cannot be counted in the numerator ($B \subset S$).
\autoref{fig:git/coverage-achievability} reveals the unachievable
coverage which other tools suggest exists but is ultimately unattainable after
accounting for each variable's defined region.

\ifonecolumn{
\begin{wrapfigure}[23]{l}{0.45\linewidth}
}{
\begin{figure}[b]
}
  \ifonecolumn{\vspace*{-2ex}}{}
  \includegraphics[trim=10 10 10 10]{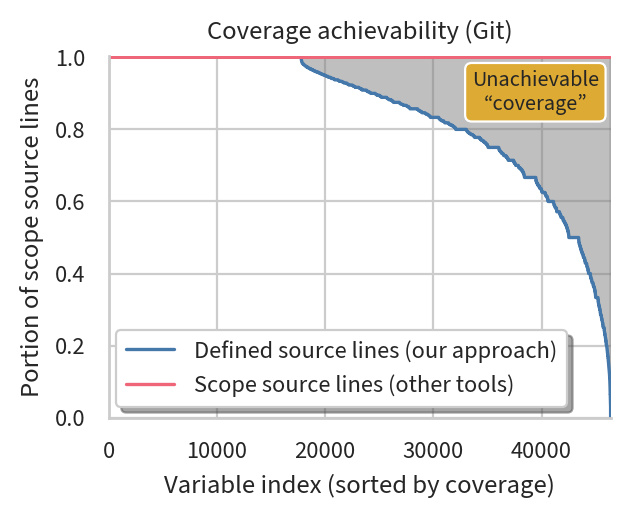}
  \caption{Coverage achievability in the Git codebase.
  The gap between the lines represents ``artificial'' coverage that
  other tools suggest exists but cannot be achieved
  as it is outside the variable's defined ranges.
  The ``other tools'' line is always 1.0 by definition
  and included only to emphasise the difference between approaches.}
  \sk{Need precise name for the coverage metric used}
  \jrs{Perhaps this should be normalised such that our approach is the constant
  1.0 value (making it match the presentation of our other coverage graphs), and
  then the other tools would appear to have excess coverage above 1.0?}
  \label{fig:git/coverage-achievability}
\ifonecolumn{
\end{wrapfigure}
}{
\end{figure}
}

% A simple attempt at fixing this latter problem would be to count in the numerator
% only those program points where the variable is defined.
%
% \[
% C_{v} =  \frac{\bigm|B_v \cap D_v \bigm|}
%               {\bigm|S_v\bigm|}        \text{~~for~~} P \text{~~the program's set of instruction bytes}
% \]
%
% This avoids penalising register allocators, but retains the problem that
% the denominator $|S_v|$ overestimates the range of
% instructions of which the debug info can possibly define a meaningful value for the variable:
% in general it includes locations before the value is defined (assigned to),
% i.e.\ program points in $S$ but not in $D$.
% This overestimation means that now both for stack- and register-allocated locals,
% full coverage becomes impossible
% to achieve, since the denominator counts program points
% over which the variable has no defined value.

\subsection{Counting Source Lines, not Instructions}

To address the first two problems---incomparability across compilers,
and omitting ``source-only'' coverage---we can count source lines, not instructions.
This applies to both the numerator and the denominator.

\vspace*{-2.0ex}
\[
C_{v} =  \frac{\bigm|B_v\bigm|}
              {\bigm|S_v\bigm|}        \text{~~for~~} P \text{~~the program's set of source lines}
\]
\vspace*{-1.0ex}

The debug info includes a line table, supplying a mapping such that
any set of source lines can be projected to a set of instructions or vice-versa.
By performing this projection, we obtain
a set of source positions (e.g.\ pairs of (filename, line number))
that should be embodied by \emph{any} compilation of the same code.
When $P$ is defined in this way,
the number of instructions embodying any particular source location
may vary arbitrarily without affecting the metric.
This is intentional: we deem
instruction counts
not relevant to the experience of a developer debugging at source level.\footnote{
The focus on lines does mean that very long source lines, e.g.\ as produced by some macro expansions,
suffer limited resolution.
However, our metric could work with any notion of \emph{program points}
that is reflected in the compiler's line table.
For example, source features could be correlated to \DWARF{}
column information, but compiler support for this information is patchy.}
%by a developer debugging at source level.
%Put differently: while the instruction count of course affects performance of the generated code,
%it is of no direct relevance to the debug coverage as experienced
%by a developer debugging at source level.

Line tables' mappings are not one-to-one:
inlining leads to source lines mapping to many program counters,
while various code-folding transformations
do the converse.
While we may still calculate sets of source lines or instruction bytes,
a local variable may be covered or not covered at each.
The simplest scheme
(which is what we implement)
calculates a variable's coverage as an unweighted mean
across all realised instances of a source line, e.g.\ inlined copies
plus the out-of-line copy.
Each source line continues to contribute at most 1 to the metric's numerator, so in the case of
such ``multiply instantiated'' lines, a contribution may be fractional
rather than just 0 or 1.\footnote{For exposition we continue to use the set cardinality notation,
in spite of this generalisation to fractional contributions.
Another fractional case might be where
only certain bytes of a value are available,
as common with locals of \code{struct} or other composite type.
Our tool does not currently account for this, but could easily be extended to do so.}
(One could also argue for unequal weighting among these instances,
e.g.\ supposing the out-of-line copy is called more often than each inlined copy.
Making a meaningful choice would require profile information.)
% In \autoref{sec:inlining}
% we discuss the potential benefits of a weighted counting approach
% rather than equal-weight counting of the resulting sets' cardinalities.)
\sk{The inlining thing needs to be written}
\jrs{Commented out reference for now, to attempt to hit page limit}

\subsection{Correcting Accidents, Permitting Full Coverage}
\label{sec:scope-shrinking}

To correct the accidental favouring of stack allocation
and unachievability of full coverage,
the obvious evolution is to apply a ``definedness filter''
to both numerator and denominator.
We call this ``scope shrinking''.
(This can be done while counting either source lines or
instruction bytes; we use only the former from here on.)
We simply omit to count, in either numerator or denominator,
those program positions for which the variable has no defined value.
\[
C_{v} =  \frac{\bigm|B_v \cap D_v \bigm|}
              {\bigm|S_v \cap D_v \bigm|}        \text{~~for~~} P \text{~~the program's set of source lines}
\]
Unlike the earlier metrics,
this is challenging to implement.
How should we calculate $D_v$,
the set of source lines in the \emph{defined range} of variable $v$?
A form of binary liveness analysis could be used,
but would require control-flow reconstruction on the binary,
and would be assuming correctness of the compiler-generated line table.

\def\sqPDF#1#2{0 0 m #1 0 l #1 #1 l 0 #1 l h}
\def\trianPDF#1#2{0 0 m #1 0 l #2 4.5 l h}
\def\uptrianPDF#1#2{#2 0 m #1 4.5 l 0 4.5 l h}
\def\circPDF#1#2{#1 0 0 #1 #2 #2 cm .1 w .5 0 m
   .5 .276 .276 .5 0 .5 c -.276 .5 -.5 .276 -.5 0 c
   -.5 -.276 -.276 -.5 0 -.5 c .276 -.5 .5 -.276 .5 0 c h}
\def\diaPDF#1#2{#2 0 m #1 #2 l #2 #1 l 0 #2 l h}

\def\credCOLOR   {.54 .14 0}
\def\cblueCOLOR  {.06 .3 .54}
\def\cgreenCOLOR {0 .54 0}
\def\cblackCOLOR {0 0 0}

\def\cpblueCOLOR {0 .45 .70}
\def\cporanCOLOR {.87 .56 .02}
\def\cpligrCOLOR {.01 .62 .45}
\def\cpdagrCOLOR {.04 .24 .18}

\def\genbox#1#2#3#4#5#6{% #1=0/1, #2=color, #3=shape, #4=raise, #5=width, #6=width/2
    \leavevmode\raise#4bp\hbox to#5bp{\vrule height#5bp depth0bp width0bp
    \pdfliteral{q .5 w \csname #2COLOR\endcsname\space RG
                       \csname #3PDF\endcsname{#5}{#6} S Q
             \ifx1#1 q \csname #2COLOR\endcsname\space rg
                       \csname #3PDF\endcsname{#5}{#6} f Q\fi}\hss}}

                                    % shape     raise width  width/2
\def\sqbox      #1#2{\genbox{#1}{#2}  {sq}       {0}   {4.5}  {2.25}}
\def\trianbox   #1#2{\genbox{#1}{#2}  {trian}    {0}   {5}    {2.5}}
\def\uptrianbox #1#2{\genbox{#1}{#2}  {uptrian}  {0}   {5}    {2.5}}
\def\circbox    #1#2{\genbox{#1}{#2}  {circ}     {0}   {5}    {2.5}}
\def\diabox     #1#2{\genbox{#1}{#2}  {dia}      {-.5} {6}    {3}}

\newcommand{\circled}[1]{\raisebox{.5pt}{\textcircled{\raisebox{-.9pt} {#1}}\hspace*{-0.1em}} }

\begin{figure*}
\begin{tabular}{ p{0.25\linewidth}r p{0.35\linewidth}|p{0.045\linewidth}|p{0.045\linewidth}|p{0.045\linewidth}|p{0.045\linewidth} }
\multicolumn{3}{c}{instructions, source positions, source code} &\multicolumn{4}{|l}{source variables}\\
             & pos               &                                        & \code{p}                         & \code{x}                         & \code{y}                           & \code{i}\\
\ifonecolumn{
\multirow{11}{*}{\begin{adjustbox}{width=1.51\linewidth,valign=top,lap={0.08\width}{-0.08\width},margin*=0em -3.2ex}\includegraphics{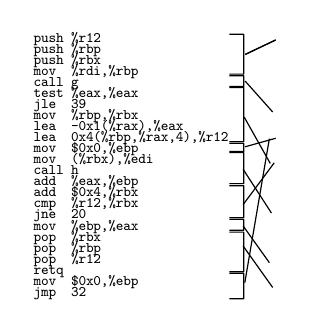}\end{adjustbox}}
}{
\multirow{11}{*}{\begin{adjustbox}{width=1.18\linewidth,valign=top,lap={0.08\width}{-0.08\width},margin*=2.8em -3.2ex}\includegraphics{figures/assembly-crop.pdf}\end{adjustbox}}
}
             & \circbox1{cblack} & \code{~1~}\verb+int f(int *p)+        & \sqbox1{cblack}~\trianbox1{cblack}&                                  &                                    &                          \\
             &                   & \code{~2~}\verb+{+                       &                                   &                                  &                                    &                          \\
             &                   & \code{~3~}\verb+  int x;+                &                                   &                                  &                                    &                          \\
             & \circbox1{cblack} & \code{~4~}\verb+  int y = g(p);+         & \sqbox1{cblack}~\trianbox1{cblack}& \sqbox0{cblack}~\circled{1}                 &  \sqbox1{cblack}~\trianbox1{cblack} &                          \\
             & \circbox1{cblack} & \code{~5~}\verb+  x = 0; +               & \sqbox1{cblack}~\trianbox1{cblack}& \sqbox1{cblack}~\trianbox1{cblack}&  \sqbox1{cblack}~\trianbox1{cblack} &                          \\
             & \circbox1{cblack} & \code{~6~}\verb^  for (int i=0; i<y; ++i)^&\sqbox1{cblack}~\trianbox1{cblack}& \sqbox1{cblack}~\trianbox1{cblack}&  \sqbox1{cblack}~\trianbox1{cblack} & \sqbox1{cblack}~\trianbox1{cblack} \\
             &                   & \code{~7~}\verb+  {+                     &                                   &                                   &                                     &                           \\
             & \circbox1{cblack} & \code{~8~}\verb^    x += h(p[i]);^       & \sqbox1{cblack}~\trianbox1{cblack}& \sqbox1{cblack}~\trianbox1{cblack}&  \sqbox1{cblack}~\trianbox1{cblack} & \sqbox1{cblack}~\trianbox1{cblack} \\
             &                   & \code{~9~}\verb+  }+                     &                                   &                                   &                                     &                           \\
             & \circbox1{cblack} & \code{10~}\verb+  return x;+             & \sqbox1{cblack}~\trianbox0{cblack}& \sqbox1{cblack}~\trianbox1{cblack}&  \sqbox1{cblack}~\trianbox0{cblack} &              \\
             & \circbox1{cblack} & \code{11~}\verb+}+                       &                                   &                                   &                                     &               \\\hline
             & \multicolumn{2}{r|}{coverage}                                & 5/6                               & 4/4                               & 4/5                                 & 2/2\\
\end{tabular}
\caption{An annotated view of how our coverage metric might view a simple compiled function.
Circles \circbox1{cblack} denote source positions identified in the debug info.
Square boxes mark those source positions for which a variable is in-scope and
undefined (\sqbox0{cblack}) or defined (\sqbox1{cblack}).
Triangles mark those for which the given source variable is covered by debug info (\trianbox1{cblack})
or not (\trianbox0{cblack});
the markers shown are typical for a lightly optimised version of the code.
\circled{1} marks a location
where variable \code{x} is not yet initialized
but, if allocated storage e.g.\ on the stack, would be erroneously counted as covered
by a naive metric.
%\circled{2} marks a location
%where the assignment can potentially be fully residualised over a zero-length range of instruction bytes,
%and therefore not reflected by a naive debug coverage metric.
}
\label{fig:coverage-annotated-variables}
\end{figure*}

\section{Implementation}
\label{sec:baseline-implementation}
\label{sec:implementation}
% To do better,
% we instead assume that an unoptimised version of the program
% is also available, either as source code or unoptimised intermediate representation.
% We use this unoptimised version to calculate the \emph{source position(s)} of first definition.
% Such a tool will not be fooled if optimisations erroneously drop some of the debug info
% covering from this point forwards,
% and could more easily accommodate branching cases.
% \sk{Ideally would show in the figure, but it doesn't capture this case}

% The metrics as described thus far are appealing because they can be computed
% from a single binary---without reference even to the source code
% (besides the binary's line table).
% However,
% %this is also the source of problems:
% we noted earlier that to calculate the denominator $|S \cap D|$
% on a binary would require a complex binary analysis.
This section outlines how our implementation avoids
a complex binary analysis and avoids assuming
correctness of the compiler's line table, by
adopting an external baseline.
%
%such as the original source program,
%produces more reliable results.

\paragraph{Heuristic attempts}
To avoid the complexity of control-flow reconstruction
%a control-flow reconstruction
%and liveness analysis,
the LLVM tool's developers trialled a modification
where, essentially, an approximation of
$|S \cap D|$
was used.
If $S$ consists of a set of instruction bytes, as offsets within a function,
the denominator is $|S \setminus [0, n)|$ where byte $n$ is the
first over which \emph{any} \DWARF{} expression for the variable is defined.
However, this is circular:
it uses what is \emph{covered}
(the instructions covered by defined \DWARF{} expressions)
as a proxy for what is \emph{coverable}
(the instructions over which the variable \emph{should} be covered),
undermining the intention of a coverage metric to capture the gap between these.
Indeed the heuristic was
found to be unreliable and removed \cite{bessonovaLlvmdwarfdumpStatisticsUnify2019}.
%coverage gaps occurring earlier in instruction space,
%over ranges where the variable's value \emph{is} in reality defined but the relevant debug info is missing,
%were being silently ignored because the heuristic conflates this with the not-yet-initialized case.

\paragraph{Need for an external baseline}
% In fact, even after implementing the complex binary analysis outlined earlier,
% the logic would remain somewhat circular, since it would be
% assuming completeness of the line table---yet compiler bugs might
% conceivably lead to source locations being omitted from that table.
% This is a generalised form of the circularity we
% just observed.
To avoid this circularity
and
%Therefore, to
answer more authoritatively which source positions \emph{should} be covered,
we can use an external reference,
in the form of
some unoptimised version of the program.
This could be either source code or unoptimised intermediate representation.
We use this unoptimised version to calculate the sets of source positions
$S$ and $D$ needed for the denominator.
%Such a tool will not be fooled if optimisations erroneously drop some of the debug info
%covering from this point forwards,
%and could more easily accommodate branching cases.
\sk{Ideally would show in the figure, but it doesn't capture this case}

\paragraph{\code{mem2reg} as a baseline}\label{sec:mem2reg}
To approximate $D$ by excluding those lines which \emph{precede
the variable's point(s) of first definition},
one method is to use the compiler itself.
Although a fully unoptimised build does not help (because
the compiler usually just places all variables on the stack
for their entire scope),
adding \emph{just} LLVM's \code{mem2reg} pass to the unoptimised \code{O0} build
largely has the right effect: it moves variables into registers
and describes them only from their point(s) of definition.
Essentially, this exploits \code{mem2reg}'s liveness analysis
to approximate $S \cap D$.
Since this pass works on virtual registers, it is not limited by the supply of physical registers,
so is applied to any non-address-taken scalar local.
% We use it only for filtering out from the baseline (denominator) those lines that precede the first definition;
% we do \emph{not} make symmetric use of
% \sk{Need to mention the assumption of textually forward control flow}
% can be used to measure the coverage of GCC or any other compiler's output.
We call this baseline \code{O0-mem2reg}.
\sk{Would enumerating lines $S_v$ be reliable from just the binary,
even if we did a bespoke binary analysis with control-flow reconstruction?
I want to argue not, but not sure.}
%\paragraph{Problems with \code{mem2reg}}
Unfortunately, various
uncontrollable effects make
\code{O0-mem2reg} unsuitable as a baseline. %the \code{O0-mem2reg} baseline
At different optimisation levels, we found that the line number tables
map the same code to slightly different source ranges
and that \code{O0-mem2reg} tends to yield a slight subset of the true $D$.
Some \code{O1} compilations
report source lines absent from the \code{O0-mem2reg} baseline set,
leading to coverage over 100\%, which is clearly
neither accurate nor acceptable.

%Both of these mean that the lines covered at \code{O1} are not a subset of those in the

\sk{epilogue/prologue anecdote might be nice here}

\paragraph{Source analysis} Instead of \code{mem2reg}, we use a source-level analysis
to calculate $S$ and $D$ as sets of source line numbers.
These sets are constructed by a series of ``filters'' over the source line number space.\sk{Is it
really a series of filters, or just one? Sorry for butchering this.}\jrs{
Well, I have thus far thought of
(and implemented, see also old steps in \code{offcuts/implementation.tex})
both the
``only computational lines'' and
``region starts after line of first definition''
steps as filters.
In your presentation here, they are treated in different ways:
filtering to computation has ``already happened''
as part of the definition of $S$ and $D$
(as mentioned a few sentences down),
while defined region is represented as its own set ($D$)
which is actually intersected in your formulas.
Perhaps as you've written here then,
only the ``computational'' part should be thought of as a filter
(which happens while computing $S$ and $D$)?}
% First,
We classify AST nodes according to whether they \emph{perform computation}---for example,
declaration-only nodes do not perform computation,
but nodes using variables and invoking operators do.
(Note that this computation may later be residualised into debug info; there is no obligation
for this computation to be embodied by \emph{instructions} in the eventual binary.)
We use the parser's retained source coordinates to map nodes to lines.
Any \emph{non-}computational lines that appear in the \DWARF{} line table
are ignored when calculating the
measured binary's set of described lines ($B$),
sidestepping the \code{O0} and \code{O1} variability
seen with \code{mem2reg}.
%This ensures the measured set will be a subset of the computational lines enumerated in the source file.
The source analysis similarly computes $S$ and $D$ as subsets of the file's computational lines.
Since $S$ and $D$ are determined by source analysis,
our denominator remains valid
even when compilers' line tables are incomplete.

% Secondly, for a given variable
% we calculate its defined region at source level (from its first line of definition to end of scope).
% When measuring coverage in  a given binary,
% lines in its \DWARF{} line table
% which are not part of
% but which
%
% We apply these filters to the raw lines covered in \DWARF{} from each variable
% (after mapping bytes covered to lines via the line table).
% Any claimed coverage outside of the intersection of these filters
% is considered incorrect and/or unimportant and thus dropped.
% We also apply the same filters to each variable's
% parent scope (block, function, etc.)
% to determine the achievable coverage
% (and our coverage ratio denominator).
\sk{See commented-out stuff here: I think this paragraph is redundant w.r.t.\ what comes before, but I'm not 100\% sure}
\jrs{You are on the right track overall, see my other replies above.}

\paragraph{Static versus dynamic analysis}\label{sec:static-analysis}\label{sec:static-versus-dynamic}\label{sec:reachability}
The approach we have described is a static analysis
examining all code; it does not rely on collecting execution traces.
However, in the presence of unreachable code
it may be fairer to filter out lines that are not executed
across some collection of test inputs.
Our method is adaptable straightforwardly to this approach,
simply by an additional filter on the sets of lines in the numerator and denominator.
We describe an experiment using this approach in the next section.

\section{Evaluation}
\label{sec:evaluation}
\label{sec:experiments}

Of our four research questions, RQ1 and RQ2 have been
addressed by construction in the preceding sections.
In this section we
first outline some practical experience
with our metric across a range of codebases and compilers/versions, then
consider the remaining questions RQ3 and RQ4.

\subsection{Experience with Our Metrics}
\label{sec:metrics}
\label{sec:metric-experiments}

\begin{figure}
  \includegraphics[trim=10 10 10 10]{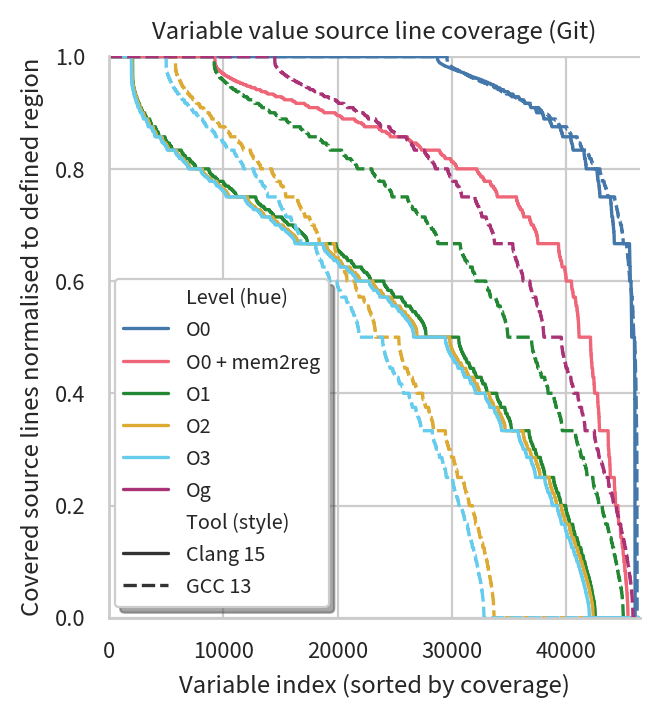}
  \caption{Variable value source line coverage for Git codebase.
  For each optimisation level, variables are sorted
  by coverage (variable's index may be different for each line).
  \sk{Need precise name for the coverage metric used}}
  \label{fig:git/coverage-ratio-levels}
\end{figure}

\autoref{fig:git/coverage-ratio-levels}
shows coverage achieved for the Git codebase as compiled by Clang 15 and GCC 13,
as measured using our metric.
Coverage is computed per variable
and plotted after sorting the variables by coverage.
Each compiler is run at multiple optimisation levels.
% (compiler version and optimisation level)
% During our evaluation, we found that coverage for Git
% as compiled by Clang and GCC generally trends towards long-term improvement,
% although Clang's \code{O1} shows a sizeable regression from Clang 12 to 15.
As one might expect,
when comparing coverage across optimisation levels,
we see that any optimisation beyond \code{O0}
significantly degrades coverage.
GCC's \code{Og} optimisation level
(which is a variant of \code{O1} tuned for better debugging)
offers the best coverage of the optimised runs.
Clang does not currently offer a meaningful \code{Og}
(it is merely an alias for \code{O1}),
but the LLVM community is working on \cite{livermore-tozerRFCRedefineOg2023}
adding this in the near future.
Recent versions of Clang enable almost all optimisations at \code{O1},
so the higher levels show only minor differences.
The difference between \code{O0} and \code{O0-mem2reg}
(which moves most variables off the stack and into registers)
is intriguing,
since it shows a sizeable amount of coverage lost
even after little optimisation.
GCC fully covers more variables than Clang
at the same setting, but its coverage then drops more rapidly.
% across the remaining variables.

\ifonecolumn{
\begin{wrapfigure}[23]{l}{0.45\linewidth}
}{
\begin{figure}
}
  \includegraphics[trim=10 10 10 10]{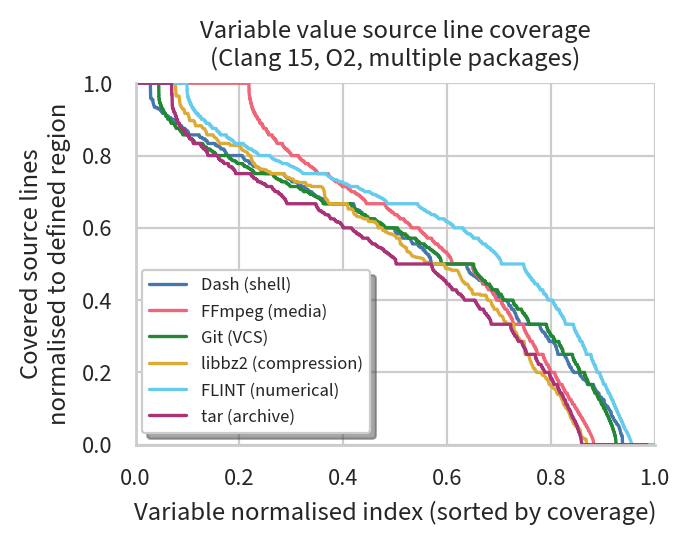}
  \caption{Variable value source line coverage across several packages.
  Variables are sorted by coverage.}
  \label{fig:coverage-ratio-packages}
\ifonecolumn{
\end{wrapfigure}
}{
\end{figure}
}

\begin{figure}[t]
  \begin{subfigure}[c]{0.49\columnwidth}
    \includegraphics[trim=10 10 10 10,scale=0.7]{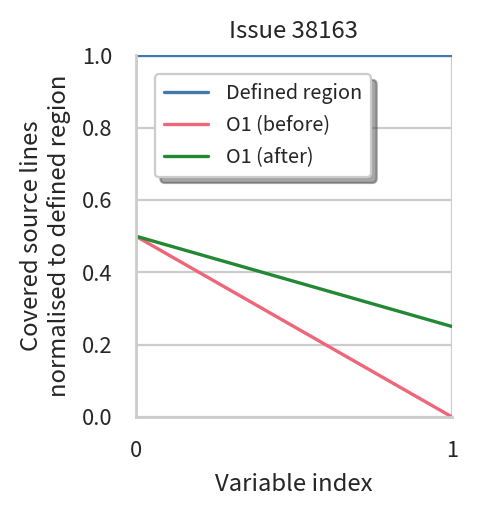}
  \end{subfigure}
  \begin{subfigure}[c]{0.49\columnwidth}
    \includegraphics[trim=10 10 10 10,scale=0.7]{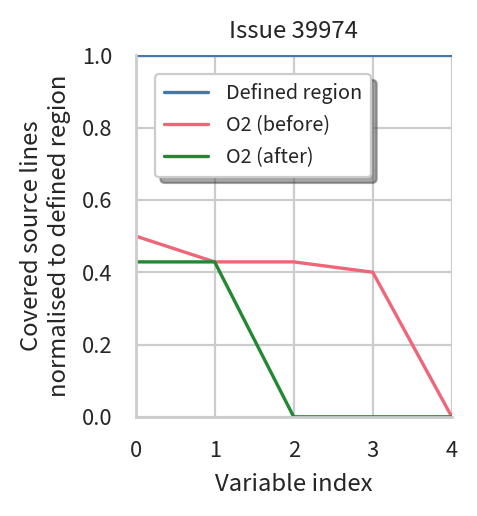}
  \end{subfigure}
  \caption{Coverage
  before and after
  resolving LLVM issues 38163 (left) and 39974 (right).
  Each measured program is the (very small) bug reproduction example
  attached to the corresponding issue,
  with only 2 and 5 variables respectively.}
  \label{fig:case-studies}
\end{figure}

\autoref{fig:coverage-ratio-packages}
shows coverage across 6 analysed packages.
These packages were chosen to explore
coverage across different application domains.
The list includes
Dash (shell),
FFmpeg (media),
Git (VCS),
libbz2 (compression),
FLINT (numerical), and
tar (archive).
%Some are ubiquitous packages (e.g.\ tar),
%while others are more niche (e.g.\ FLINT).
%Two packages (libbz2, FLINT) are libraries, while the rest are applications.
Compiled with Clang 15 at \code{O2},
we see a similar profile across all packages.

We compared our coverage metric approach
with that of other tools
(\code{llvm-dwarfdump}, \code{debuginfo-quality})
for the Git codebase using Clang 15 at \code{O2}
by computing the correlation between the two.
The Pearson correlation coefficient is
0.656,
which suggests only a moderately correlated positive relationship.
Indeed, our approach often computes quite different coverage values
than past approaches \cite{llvmprojectLlvmdwarfdumpDumpVerify2020,
ocallahanComparingQualityDebug2018}.
Several factors explain this.
As highlighted earlier (\autoref{sec:illusion-coverage-defined}),
we lift from bytes to lines, restrict to
computational lines, and include only the defined region
as covered and coverable.
These improvements affect both
numerator and denominator, and manifest differently for each variable
depending on whether each line of its defined region
is covered or uncovered.

\subsection{Case Studies}
\label{sec:case-studies}

RQ4 asked: in detail, does our metric explain debuggability gaps
in a way consistent with how these are understood by real compiler developers?
We examine this through two case studies.
Compiler changes can go both ways:
even when an issue is \emph{resolved}, coverage may go up or down.
This is because compiler authors generally take the
(quite reasonable) perspective that incorrect debug info is worse than none at
all \cite{llvmprojectSourceLevelDebugging2022},
so may sometimes remove coverage to avoid incorrectness.
We study one case of each kind.

\paragraph{Debug info repaired}
LLVM issue 38163 (``Loop strength reduction preserves little debug info'')
involves the \code{LoopStrengthReduce} pass which transforms loop induction
variables into more efficient forms. Debug info describing these
induction variables was being dropped.
The issue was resolved via a change which residualises the variable in some cases.
% Our consistency checking approach detected this issue and verified the fix.
As seen in \autoref{fig:case-studies}, our metric shows that
source variable value coverage improved as a result.

\paragraph{Debug info dropped}
LLVM issue 39974 (``Salvaged memory loads can observe subsequent memory
writes'') concerns a function that loads from memory
into an unused local variable. The \code{EarlyCSE} pass eliminates
the load but correctly residualises this as a dereference operation
in debug info. However, after a later store to the pointer
this expression yields an incorrect value: the newly
stored value, not the value that would have been loaded earlier
(which has become potentially unknowable).
The compiler authors chose to remove the debug info for this variable.
% Our consistency checking approach detected this issue and verified that info was
% removed as intended after resolution.
\autoref{fig:case-studies} shows that our metric confirms that
several source variables lost all value coverage as expected for this compiler
change.
\sk{Extra knowability discussion could go here}

\subsection{Aggregate Comparisons: A Replication Study}
\label{sec:replication}

\sk{I am not doing justice to the fact that the debuginfo-quality
comarisons have partially addressed RQ3 already. This text would ideally be
refactored a bit to show this.}

RQ3 asked: in aggregate, how does our metric's picture of debuggability
depart from those computed by na\"ive variations and/or previously proposed metrics?
We answer this
%We will answer this open question
%both by applying our metric to both real and synthetic programs
%and exploring the results in plots,
%and
by a \emph{replication study}
in which we adapt our metrics to reproduce
an experiment from recent literature.
The similarities we find serve as validation of definitional and implementational
correctness
but we also find certain expected
differences that we can explain in terms of our metric's definition.
The study reproduces the experiments of Assaiante et al.\ \cite{assaianteWhereDidMy2023}
which generated the data plotted in \autoref{fig:replication/comparison}.
The metrics presented have been computed across
the same 5,000 Csmith-generated programs
from their published artifact.
Their metrics examine local variables only, not formal parameters,
so we created a similarly adapted version of ours.
% (rather than both locals and parameters),

\begin{figure*}
  \includegraphics[trim=10 10 10 0,scale=0.98]{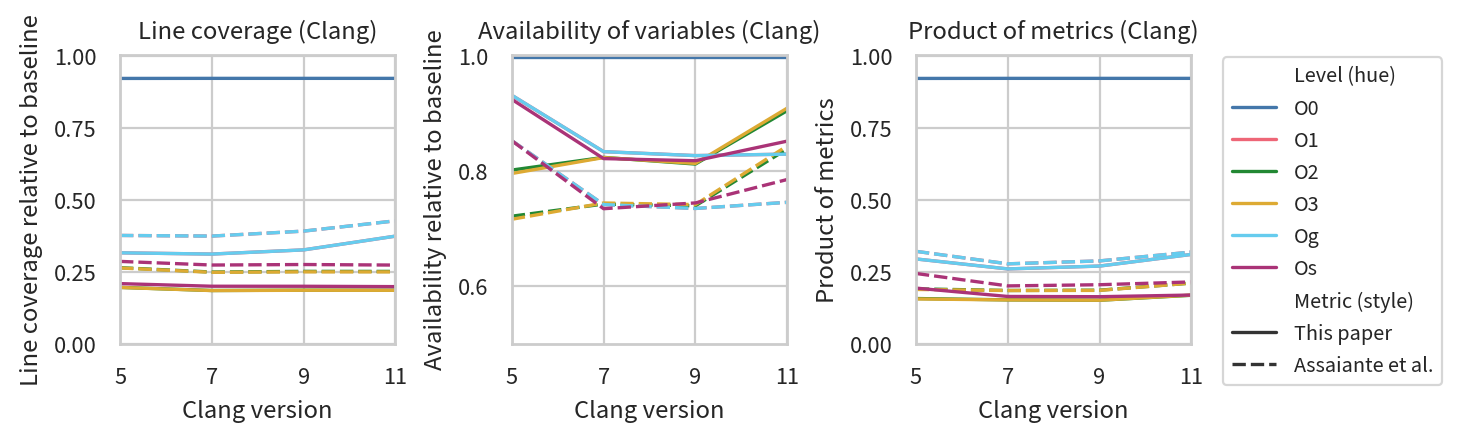}
  \caption{Comparison of our metric approach
  (using static source analysis as baseline)
  with Assaiante et al.\ \cite{assaianteWhereDidMy2023}
  (using \code{O0} as baseline).
  Metrics computed using the same 5,000
  Csmith-generated \cite{yangFindingUnderstandingBugs2011}
  programs as used by Assaiante et al.
  Of the two, only our metric is capable of measuring coverage at \code{O0}.
  The two metrics show the same trends across optimisation levels and versions
  but a markedly different absolute value, reflecting our
  fairer ``scope shrinking'' baseline (\autoref{sec:scope-shrinking}).
%  Line coverage gap between metrics is expected due to use of different baselines.
%  Availability of variables difference is due to key strength of only our metric:
%  we reduce each variable's scope such that
%  it starts only from its first definition.
}
  \label{fig:replication/comparison}
\end{figure*}

%Their work uses \code{O0} as a baseline
%and computes an overall summary in a ``line-focused'' manner.
%they also present the ``product of metrics'',
%which weights the availability score by lines present.
%We have replicated their values ourselves
%by re-running their experiments
%and confirming them with the same figure from their work.
\sk{This was unclear. What did we show? I have tried
to answer this above with `The similarities we find\ldots{}' but it
needs to be spelled out more clearly.}

While we have defined our own metric in a per-variable fashion,
aggregating across lines,
Assaiante et al.\ did the converse,
defining a metric per line and aggregating over that line's variables.
They first compute ``line coverage'' as a ratio of
the lines present in the \DWARF{} line table for a given compilation
relative to the lines present in their \code{O0} baseline.
Next, they compute ``availability of variables''
by attempting to stop in the debugger on each such line,
and recording the number of variables
accessible relative to \code{O0}.
The overall availability score is the arithmetic mean.
Fortunately, our method
is straightforwardly adapted to a line-oriented approach,
simply by enumerating at each line the familiar denominator
$|S \cap D|$ (the variables that are both in-scope and defined)
and numerator $|B \cap D|$ (the variables that are both described and defined).
Our approach is able to compute
the metric for \code{O0},
since we use our familiar source analysis as a baseline,
whereas Assaiante et al.\ use \code{O0} as the baseline;
we noted in \autoref{sec:mem2reg} why this is unreliable.
\sk{Please check what I wrote.
And did we find a difference of the kind expected? Why or why not?}

% and risks missing some coverage.
% While the overall trend is towards fewer  lines covered
% as optimisation increases,
% we found that compilers do not strictly reduce lines:
% instead, the next optimisation may attribute instructions
% to a slightly different source line
% which may not have appeared in \code{O0}.
% Of the two approaches,
% only our metric can tolerate these line shifts.

%elsewhere in the paper,
%here we switch to the same ``line-focused'' approach

Their work relies on a dynamic analysis
%which attaches to the executing program in the debugger
%and tries to observe variables at run time,
and thus will not encounter
unreachable lines,
which these generated programs do contain.
For this replication experiment,
we added a binary analysis step
using a simple custom Valgrind \cite{nethercoteValgrindFrameworkHeavyweight2007b} tool
to enumerate executed lines,
as anticipated earlier (\autoref{sec:static-versus-dynamic}).
This allowed us to obtain similar line coverage.

Similar trends appear in both our metric and theirs
when looking across compiler versions and optimisation levels.
Our line coverage is slightly lower
owing to the different baseline.
Since our metric can measure line coverage even at \code{O0},
we can additionally see that if their \code{On} coverage
is multiplied by the \code{O0} value (which is their baseline),
we indeed arrive at (approximately) our own \code{On} line coverage.
Meanwhile, our availability of variables
is somewhat higher.
This reflects the expected improvement of our approach:
by limiting coverable lines to the defined region,
we avoid the problem of the over-counting stack-based \code{O0} baseline,
which artificially prevents reaching 100\% coverage (\autoref{sec:register-allocation}).
%where variables are on the stack and
%appear available even where they are undefined.
%Owing to the register allocation issue identified earlier
%
%doing so
%artificially prevents

\ifonecolumn{
\begin{wrapfigure}[23]{l}{0.45\linewidth}
}{
\begin{figure}
}
  \includegraphics[trim=10 10 10 10]{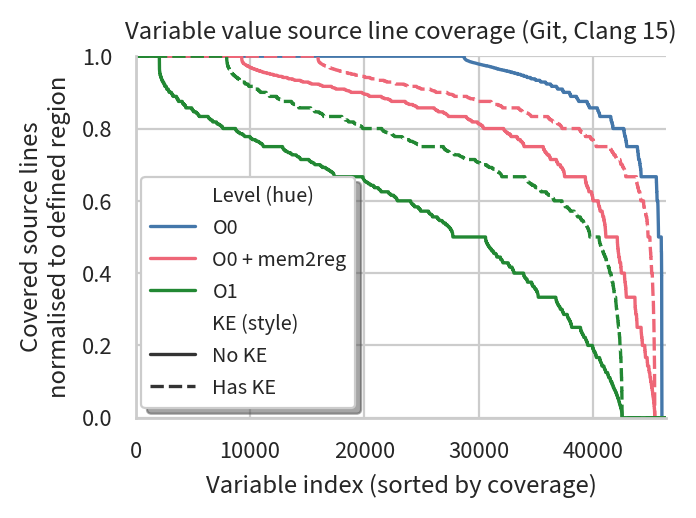}
  \caption{Coverage for the Git codebase before and after end-of-scope
  ``knowledge extension'' (KE).}
  \label{fig:git/coverage-ratio-knowledge-extension}
\ifonecolumn{
\end{wrapfigure}
}{
\end{figure}
}
\vspace*{-2ex}

\subsection{Experimenting with End-Of-Scope Treatments}
\label{sec:knowability-extension}
\label{sec:knowledge-extension}
Scope shrinking (\autoref{sec:scope-shrinking})
disregards a variable's region of undefinedness at the beginning
of its scope, i.e.\ until it is defined.
A near-dual scenario exists at the \emph{end} of its scope:
a defined variable may have a period of \emph{unknowability}
when its last value, although well-defined in source terms,
is no longer recoverable from the object program state (seen in \autoref{fig:liveness-diagram}).
Since \DWARF{} expressions residualise computation,
but not  state, even the most sophisticated \DWARF{} expression could not reconstruct the variable in this case.
However, a hypothetical debugger could be implemented today
that performs this \emph{knowledge extension}, by saving a variable's value
at its last moment of knowability (computed from the instruction ranges of debug info).
\autoref{fig:git/coverage-ratio-knowledge-extension} shows a simulation\todo{
Actually implement this, then stop calling it simulation}
of the potential coverage gains after applying end-of-scope knowledge
extension.
We do not know of any such debugger, although
\DWARF{} version 5 alludes to such possibilities
(in the specification of \code{DW\_OP\_entry\_value},
the operation to obtain \emph{somehow} a value as it existed
on entry to the function)
and
\emph{omniscient debuggers} (e.g.\ Pernosco)
take the extreme position of ensuring full knowability at all times.

\section{Related Work}
To our knowledge,
relatively little work has addressed the problem
of measuring debug info coverage.
As covered earlier
(\autoref{sec:instruction-based})
two existing tools
\code{llvm-dwarfdump}
\cite{llvmprojectLlvmdwarfdumpDumpVerify2020}
and \code{debuginfo-quality}
\cite{ocallahanComparingQualityDebug2018}
compute
essentially the same instruction-based metric,
whose problems we surveyed.
\citet{assaianteWhereDidMy2023} used a differential approach which
tracks how many variables can be accessed in the debugger on each source line
relative to the unoptimised version.
Our metric offers more precision via our source analysis baseline and
by considering only each variable's defined region.
%we discussed this in \autoref{sec:replication}.

Rather more work has addressed
other aspects of source-level debugging of optimised code.
Research into debugging optimised programs made some progress in the 1980s
\cite{hennessySymbolicDebuggingOptimized1982}, mainly focused on tracking the
location of source variable values in optimised program state during execution.
DOC \cite{coutantDOCPracticalApproach1988} stores debug info
tracking locations of values from memory through to registers,
resembling an early form of today's debug info.
\citet{zellwegerInteractiveSourcelevelDebugging1984} describes
how to compare unoptimised and optimised program data-flow and
control-flow to untangle state changes of specific optimisations and recover
% variable
values.
OPTVIEW \citep{ticeOPTVIEWNewApproach1998} avoids
attempting to map between source and binary, but instead lifts the optimised
program back up to a modified source view.
Still other systems achieve source-level (``expected'')
behaviour by temporarily switching
to an unoptimised version of function(s) while a debugger is attached
\cite{zurawskiDebuggingOptimizedCode1991, holzleDebuggingOptimizedCode1992},
including also performing the switch mid-function using on-stack replacement
\cite{
  holzleDebuggingOptimizedCode1992,
  guoDebuggingTimeJITs2014,
  deliaOnstackReplacementDistilled2018}.
These systems sidestep the issues
motivating our work, but rely on just-in-time compilers, and
also constrain the compiler to retain specific program points where it is safe
to jump from optimised to unoptimised mode for debugging.
By contrast, ahead-of-time toolchain compilers are traditionally implemented
without such constraints, motivating our work.

We also discussed earlier (\autoref{sec:knowledge-extension}) how
a debugger could currently perform ``knowledge extension''
and thereby improve the debugging illusion for any given metadata.
Although we know no debugger doing this per se,
\emph{omniscient debuggers} such as Pernosco,
built on record/replay systems such as \code{rr} \citep{ocallahanEngineeringRecordReplay2017},
offer an extreme case: knowledge is kept indefinitely.
Recording must be enabled at the start of execution,
and on current hardware it must also narrow the envelope of executions e.g.\ by serialization
of multithreaded code.
% As a result, these techniques are less suited to in-the-field debugging
% (when it was not anticipated that a debugger would be needed)
% and are arguably less ``truthful'' than a conventional debugger such as \code{gdb}
% over \code{ptrace}, which does not bring such constraints.

\section{Conclusions and Future Work}

We have defined various properties of correct debug info,
characterising debug info
as a ``residualisation'' of the optimised-away parts of the original program.
We have elaborated these properties into coverage criteria that we have
embodied in an implemented tool
and found useful in finding and explaining real debuggability problems.

The viewpoint of ``residualisation'' is one we find especially useful.
It suggests two specific areas of
future work:
\emph{designing for state retention}
\label{sec:recovery-retention}
and \emph{designing for residualisation}.

The first is to allow a conventional debugger to ``residualise state'',
i.e.\ to preserve local variables over ranges when they would otherwise be \emph{unknowable},
as we developed in \autoref{sec:knowledge-extension}.
%keeping them available in the debugger until they go out of scope.
Although
%ot currently implemented,
today's debug info already makes this possible,
%for example by setting a ``hidden breakpoint'' on all local variables' end-of-range locations
%and snapshotting the ``last seen value'' when these breakpoints are hit.
doing it reliably would require control-flow reconstruction in the debugger
(since the numerically ``last'' instruction in an address range might not be the last to execute)
and ideally a faster breakpoint mechanism
(such as that of \citet{kesslerFastBreakpointsDesign1990}),
to avoid the slowdown of frequent traps.
%rather than the usual hardware trap approach,
%since frequent traps would cause major slowdown.

The second goal is open-ended, beginning by
observing how compilers have not been engineered from the viewpoint of
``optimisation as residualisation'' but rather ``optimisation as elimination''.
The generation of debug info is an additional manual task for the author of
an optimisation pass.
Could it instead be an automatic side-effect of the primitive IR transformations
provided by a compiler framework?
This stands to align
debug info generation with verified compilation
\citep{leroyFormalVerificationRealistic2009}
and translation validation
\citep{neculaTranslationValidationOptimizing2000,
lopesAlive2BoundedTranslation2021}.

\begin{acks}
  We thank
Maxine Champion,
Al Grant,
Robert O'Callahan,
and
the anonymous reviewers
for their helpful feedback.
This work was supported by the \grantsponsor{EPSRC}{Engineering and Physical
Sciences Research Council (EPSRC)}{https://www.ukri.org/councils/epsrc/} via
grant \grantnum{EPSRC}{EP/W012308/1}.

\end{acks}

\balance
\bibliography{bibliography}


\begin{thebibliography}{34}

%%% the Web address.

\ifx \showCODEN    \undefined \def \showCODEN     #1{\unskip}     \fi
\ifx \showDOI      \undefined \def \showDOI       #1{#1}\fi
\ifx \showISBNx    \undefined \def \showISBNx     #1{\unskip}     \fi
\ifx \showISBNxiii \undefined \def \showISBNxiii  #1{\unskip}     \fi
\ifx \showISSN     \undefined \def \showISSN      #1{\unskip}     \fi
\ifx \showLCCN     \undefined \def \showLCCN      #1{\unskip}     \fi
\ifx \shownote     \undefined \def \shownote      #1{#1}          \fi
\ifx \showarticletitle \undefined \def \showarticletitle #1{#1}   \fi
\ifx \showURL      \undefined \def \showURL       {\relax}        \fi
% invisible to TeX
\providecommand\bibfield[2]{#2}
\providecommand\bibinfo[2]{#2}
\providecommand\natexlab[1]{#1}
\providecommand\showeprint[2][]{arXiv:#2}

\bibitem[Assaiante et~al\mbox{.}(2023)]%
        {assaianteWhereDidMy2023}
\bibfield{author}{\bibinfo{person}{Cristian Assaiante},
  \bibinfo{person}{Daniele~Cono D'Elia}, \bibinfo{person}{Giuseppe~Antonio
  Di~Luna}, {and} \bibinfo{person}{Leonardo Querzoni}.}
  \bibinfo{year}{2023}\natexlab{}.
\newblock \showarticletitle{Where Did My Variable Go? {{Poking}} Holes in
  Incomplete Debug Information}. In \bibinfo{booktitle}{\emph{Proc. of
  {{ASPLOS}} '23}}.
\newblock
\urldef\tempurl%
\url{https://doi.org/10.1145/3575693.3575720}
\showDOI{\tempurl}


\bibitem[Bessonova(2019)]%
        {bessonovaLlvmdwarfdumpStatisticsUnify2019}
\bibfield{author}{\bibinfo{person}{Kristina Bessonova}.}
  \bibinfo{year}{2019}\natexlab{}.
\newblock \bibinfo{title}{[llvm-dwarfdump][{{Statistics}}] {{Unify}} Coverage
  Statistic Computation}.
\newblock
\newblock
\urldef\tempurl%
\url{https://reviews.llvm.org/D70548}
\showURL{%
\tempurl}


\bibitem[Brender et~al\mbox{.}(1998)]%
        {brenderDebuggingOptimizedCode1998}
\bibfield{author}{\bibinfo{person}{Ronald~F. Brender},
  \bibinfo{person}{Jeffrey~E. Nelson}, {and} \bibinfo{person}{Mark~E.
  Arsenault}.} \bibinfo{year}{1998}\natexlab{}.
\newblock \showarticletitle{Debugging Optimized Code: {{Concepts}} and
  Implementation on {{DIGITAL Alpha}} Systems}.
\newblock \bibinfo{journal}{\emph{Digital Technical Journal}}
  \bibinfo{volume}{10}, \bibinfo{number}{1} (\bibinfo{year}{1998}),
  \bibinfo{pages}{81--99}.
\newblock


\bibitem[Brooks et~al\mbox{.}(1992)]%
        {brooksNewApproachDebugging1992}
\bibfield{author}{\bibinfo{person}{Gary Brooks}, \bibinfo{person}{Gilbert~J.
  Hansen}, {and} \bibinfo{person}{Steve Simmons}.}
  \bibinfo{year}{1992}\natexlab{}.
\newblock \showarticletitle{A New Approach to Debugging Optimized Code}. In
  \bibinfo{booktitle}{\emph{Proc. of PLDI '92}}.
\newblock
\urldef\tempurl%
\url{https://doi.org/10.1145/143095.143108}
\showDOI{\tempurl}


\bibitem[Coutant et~al\mbox{.}(1988)]%
        {coutantDOCPracticalApproach1988}
\bibfield{author}{\bibinfo{person}{Deborah~S. Coutant}, \bibinfo{person}{Sue
  Meloy}, {and} \bibinfo{person}{Michelle Ruscetta}.}
  \bibinfo{year}{1988}\natexlab{}.
\newblock \showarticletitle{{{DOC}}: A Practical Approach to Source-Level
  Debugging of Globally Optimized Code}. In \bibinfo{booktitle}{\emph{Proc. of
  {{PLDI}} '88}}.
\newblock
\urldef\tempurl%
\url{https://doi.org/10.1145/53990.54003}
\showDOI{\tempurl}


\bibitem[D'Elia and Demetrescu(2018)]%
        {deliaOnstackReplacementDistilled2018}
\bibfield{author}{\bibinfo{person}{Daniele~Cono D'Elia} {and}
  \bibinfo{person}{Camil Demetrescu}.} \bibinfo{year}{2018}\natexlab{}.
\newblock \showarticletitle{On-Stack Replacement, Distilled}. In
  \bibinfo{booktitle}{\emph{Proc. of {{PLDI}} '18}}.
\newblock
\urldef\tempurl%
\url{https://doi.org/10.1145/3192366.3192396}
\showDOI{\tempurl}


\bibitem[Di~Luna et~al\mbox{.}(2021)]%
        {dilunaWhoDebuggingDebuggers2021}
\bibfield{author}{\bibinfo{person}{Giuseppe~Antonio Di~Luna},
  \bibinfo{person}{Davide Italiano}, \bibinfo{person}{Luca Massarelli},
  \bibinfo{person}{Sebastian {\"O}sterlund}, \bibinfo{person}{Cristiano
  Giuffrida}, {and} \bibinfo{person}{Leonardo Querzoni}.}
  \bibinfo{year}{2021}\natexlab{}.
\newblock \showarticletitle{Who's Debugging the Debuggers? {{Exposing}} Debug
  Information Bugs in Optimized Binaries}. In \bibinfo{booktitle}{\emph{Proc.
  of ASPLOS '21}}.
\newblock
\urldef\tempurl%
\url{https://doi.org/10.1145/3445814.3446695}
\showDOI{\tempurl}


\bibitem[{DWARF Debugging Information Format Committee}(2017)]%
        {dwarfDebuggingInformation2017}
\bibfield{author}{\bibinfo{person}{{DWARF Debugging Information Format
  Committee}}.} \bibinfo{year}{2017}\natexlab{}.
\newblock \bibinfo{title}{{{DWARF}} Debugging Information Format: Version 5}.
\newblock
\newblock
\urldef\tempurl%
\url{https://dwarfstd.org/doc/DWARF5.pdf}
\showURL{%
\tempurl}


\bibitem[Eigler(2006)]%
        {eiglerSystemTap2006}
\bibfield{author}{\bibinfo{person}{Frank~{Ch.} Eigler}.}
  \bibinfo{year}{2006}\natexlab{}.
\newblock \showarticletitle{Problem Solving With {SystemTap}}. In
  \bibinfo{booktitle}{\emph{Proceedings of the Linux Symposium}} (Ottawa,
  Canada).
\newblock


\bibitem[Guo(2014)]%
        {guoDebuggingTimeJITs2014}
\bibfield{author}{\bibinfo{person}{Shu-Yu Guo}.}
  \bibinfo{year}{2014}\natexlab{}.
\newblock \bibinfo{title}{Debugging in the Time of {{JITs}}}.
\newblock
\newblock
\urldef\tempurl%
\url{https://rfrn.org/~shu/2014/05/14/debugging-in-the-time-of-jits.html}
\showURL{%
\tempurl}


\bibitem[Hennessy(1982)]%
        {hennessySymbolicDebuggingOptimized1982}
\bibfield{author}{\bibinfo{person}{John Hennessy}.}
  \bibinfo{year}{1982}\natexlab{}.
\newblock \showarticletitle{Symbolic Debugging of Optimized Code}.
\newblock \bibinfo{journal}{\emph{TOPLAS}} \bibinfo{volume}{4},
  \bibinfo{number}{3} (\bibinfo{date}{July} \bibinfo{year}{1982}),
  \bibinfo{pages}{323--344}.
\newblock
\urldef\tempurl%
\url{https://doi.org/10.1145/357172.357173}
\showDOI{\tempurl}


\bibitem[H{\"o}lzle et~al\mbox{.}(1992)]%
        {holzleDebuggingOptimizedCode1992}
\bibfield{author}{\bibinfo{person}{Urs H{\"o}lzle}, \bibinfo{person}{Craig
  Chambers}, {and} \bibinfo{person}{David Ungar}.}
  \bibinfo{year}{1992}\natexlab{}.
\newblock \showarticletitle{Debugging Optimized Code with Dynamic
  Deoptimization}. In \bibinfo{booktitle}{\emph{Proc. of {{PLDI}} '92}}.
\newblock
\urldef\tempurl%
\url{https://doi.org/10.1145/143095.143114}
\showDOI{\tempurl}


\bibitem[Jel{\'i}nek(2010)]%
        {jelinekImprovingDebugInfo2010}
\bibfield{author}{\bibinfo{person}{Jakub Jel{\'i}nek}.}
  \bibinfo{year}{2010}\natexlab{}.
\newblock \showarticletitle{Improving Debug Info for Optimized Away
  Parameters}. In \bibinfo{booktitle}{\emph{{{GCC Summit}}}}.
\newblock
\urldef\tempurl%
\url{https://gcc.gnu.org/wiki/summit2010?action=AttachFile&do=view&target=jelinek.pdf}
\showURL{%
\tempurl}


\bibitem[Kessler(1990)]%
        {kesslerFastBreakpointsDesign1990}
\bibfield{author}{\bibinfo{person}{Peter~B. Kessler}.}
  \bibinfo{year}{1990}\natexlab{}.
\newblock \showarticletitle{Fast Breakpoints: Design and Implementation}. In
  \bibinfo{booktitle}{\emph{Proc. of {{PLDI}} '90}}.
\newblock
\urldef\tempurl%
\url{https://doi.org/10.1145/93542.93555}
\showDOI{\tempurl}


\bibitem[Leroy(2009)]%
        {leroyFormalVerificationRealistic2009}
\bibfield{author}{\bibinfo{person}{Xavier Leroy}.}
  \bibinfo{year}{2009}\natexlab{}.
\newblock \showarticletitle{Formal Verification of a Realistic Compiler}.
\newblock \bibinfo{journal}{\emph{Commun. ACM}} \bibinfo{volume}{52},
  \bibinfo{number}{7} (\bibinfo{date}{July} \bibinfo{year}{2009}),
  \bibinfo{pages}{107--115}.
\newblock
\urldef\tempurl%
\url{https://doi.org/10.1145/1538788.1538814}
\showURL{%
\tempurl}


\bibitem[Li et~al\mbox{.}(2020)]%
        {liDebugInformationValidation2020}
\bibfield{author}{\bibinfo{person}{Yuanbo Li}, \bibinfo{person}{Shuo Ding},
  \bibinfo{person}{Qirun Zhang}, {and} \bibinfo{person}{Davide Italiano}.}
  \bibinfo{year}{2020}\natexlab{}.
\newblock \showarticletitle{Debug Information Validation for Optimized Code}.
  In \bibinfo{booktitle}{\emph{Proc. of PLDI '20}}.
\newblock
\urldef\tempurl%
\url{https://doi.org/10.1145/3385412.3386020}
\showDOI{\tempurl}


\bibitem[{Livermore-Tozer}(2023)]%
        {livermore-tozerRFCRedefineOg2023}
\bibfield{author}{\bibinfo{person}{Stephen {Livermore-Tozer}}.}
  \bibinfo{year}{2023}\natexlab{}.
\newblock \bibinfo{title}{[{{RFC}}] {{Redefine Og}}/{{O1}} and Add a New Level
  of {{Og}}}.
\newblock
\newblock
\urldef\tempurl%
\url{https://discourse.llvm.org/t/rfc-redefine-og-o1-and-add-a-new-level-of-og/72850}
\showURL{%
\tempurl}


\bibitem[{LLVM Project}(2020)]%
        {llvmprojectLlvmdwarfdumpDumpVerify2020}
\bibfield{author}{\bibinfo{person}{{LLVM Project}}.}
  \bibinfo{year}{2020}\natexlab{}.
\newblock \bibinfo{title}{llvm-dwarfdump - {{Dump}} and Verify {{DWARF}} Debug
  Information}.
\newblock
\newblock
\urldef\tempurl%
\url{https://llvm.org/docs/CommandGuide/llvm-dwarfdump.html}
\showURL{%
\tempurl}


\bibitem[{LLVM Project}(2022)]%
        {llvmprojectSourceLevelDebugging2022}
\bibfield{author}{\bibinfo{person}{{LLVM Project}}.}
  \bibinfo{year}{2022}\natexlab{}.
\newblock \bibinfo{title}{Source Level Debugging with {{LLVM}}}.
\newblock
\newblock
\urldef\tempurl%
\url{https://llvm.org/docs/SourceLevelDebugging.html}
\showURL{%
\tempurl}


\bibitem[Lopes et~al\mbox{.}(2021)]%
        {lopesAlive2BoundedTranslation2021}
\bibfield{author}{\bibinfo{person}{Nuno~P. Lopes}, \bibinfo{person}{Juneyoung
  Lee}, \bibinfo{person}{Chung-Kil Hur}, \bibinfo{person}{Zhengyang Liu}, {and}
  \bibinfo{person}{John Regehr}.} \bibinfo{year}{2021}\natexlab{}.
\newblock \showarticletitle{Alive2: {{Bounded}} Translation Validation for
  {{LLVM}}}. In \bibinfo{booktitle}{\emph{Proc. of {{PLDI}} '21}}.
\newblock
\urldef\tempurl%
\url{https://doi.org/10.1145/3453483.3454030}
\showDOI{\tempurl}


\bibitem[Necula(2000)]%
        {neculaTranslationValidationOptimizing2000}
\bibfield{author}{\bibinfo{person}{George~C. Necula}.}
  \bibinfo{year}{2000}\natexlab{}.
\newblock \showarticletitle{Translation Validation for an Optimizing Compiler}.
  In \bibinfo{booktitle}{\emph{Proc. of {{PLDI}} '00}}.
\newblock
\urldef\tempurl%
\url{https://doi.org/10.1145/349299.349314}
\showDOI{\tempurl}


\bibitem[Nethercote and Seward(2007)]%
        {nethercoteValgrindFrameworkHeavyweight2007b}
\bibfield{author}{\bibinfo{person}{Nicholas Nethercote} {and}
  \bibinfo{person}{Julian Seward}.} \bibinfo{year}{2007}\natexlab{}.
\newblock \showarticletitle{Valgrind: {{A}} Framework for Heavyweight Dynamic
  Binary Instrumentation}. In \bibinfo{booktitle}{\emph{Proc. of {{PLDI}}
  '08}}.
\newblock
\urldef\tempurl%
\url{https://doi.org/10.1145/1250734.1250746}
\showDOI{\tempurl}


\bibitem[O'Callahan(2018)]%
        {ocallahanComparingQualityDebug2018}
\bibfield{author}{\bibinfo{person}{Robert O'Callahan}.}
  \bibinfo{year}{2018}\natexlab{}.
\newblock \bibinfo{title}{Comparing the Quality of Debug Information Produced
  by {{Clang}} and {{GCC}}}.
\newblock
\newblock
\urldef\tempurl%
\url{https://robert.ocallahan.org/2018/11/comparing-quality-of-debug-information.html}
\showURL{%
\tempurl}


\bibitem[O'Callahan et~al\mbox{.}(2017)]%
        {ocallahanEngineeringRecordReplay2017}
\bibfield{author}{\bibinfo{person}{Robert O'Callahan}, \bibinfo{person}{Chris
  Jones}, \bibinfo{person}{Nathan Froyd}, \bibinfo{person}{Kyle Huey},
  \bibinfo{person}{Albert Noll}, {and} \bibinfo{person}{Nimrod Partush}.}
  \bibinfo{year}{2017}\natexlab{}.
\newblock \showarticletitle{Engineering Record and Replay for Deployability}.
  In \bibinfo{booktitle}{\emph{Proc. of {{USENIX ATC}} '17}}.
\newblock
\urldef\tempurl%
\url{https://www.usenix.org/conference/atc17/technical-sessions/presentation/ocallahan}
\showURL{%
\tempurl}


\bibitem[Oliva(2010)]%
        {olivaConsistentViewsRecommended2010}
\bibfield{author}{\bibinfo{person}{Alexandre Oliva}.}
  \bibinfo{year}{2010}\natexlab{}.
\newblock \showarticletitle{Consistent Views at Recommended Breakpoints}. In
  \bibinfo{booktitle}{\emph{Proc. of {{GCC Summit}}}}. \bibinfo{pages}{6}.
\newblock
\urldef\tempurl%
\url{https://www.fsfla.org/~lxoliva/papers/sfn/gcc2010.pdf}
\showURL{%
\tempurl}


\bibitem[Oliva(2017a)]%
        {olivaLocationViewNumbering2017}
\bibfield{author}{\bibinfo{person}{Alexandre Oliva}.}
  \bibinfo{year}{2017}\natexlab{a}.
\newblock \bibinfo{title}{Location View Numbering}.
\newblock
\newblock
\urldef\tempurl%
\url{https://dwarfstd.org/ShowIssue.php?issue=170427.1}
\showURL{%
\tempurl}


\bibitem[Oliva(2017b)]%
        {olivaStatementFrontierNotes2017}
\bibfield{author}{\bibinfo{person}{Alexandre Oliva}.}
  \bibinfo{year}{2017}\natexlab{b}.
\newblock \bibinfo{title}{Statement Frontier Notes and Location Views}.
\newblock
\newblock
\urldef\tempurl%
\url{https://developers.redhat.com/blog/2017/07/11/statement-frontier-notes-and-location-views}
\showURL{%
\tempurl}


\bibitem[Oliva(2019)]%
        {olivaGCCGOlogyStudying2019}
\bibfield{author}{\bibinfo{person}{Alexandre Oliva}.}
  \bibinfo{year}{2019}\natexlab{}.
\newblock \bibinfo{title}{{{GCC gOlogy}}: Studying the Impact of Optimizations
  on Debugging}.
\newblock
\newblock
\urldef\tempurl%
\url{https://www.fsfla.org/~lxoliva/writeups/gOlogy/gOlogy.html}
\showURL{%
\tempurl}


\bibitem[Stinnett and Kell(2024)]%
        {debugInfoMetricsArtifact}
\bibfield{author}{\bibinfo{person}{J.~Ryan Stinnett} {and}
  \bibinfo{person}{Stephen Kell}.} \bibinfo{year}{2024}\natexlab{}.
\newblock \bibinfo{title}{Accurate Coverage Metrics for Compiler-Generated
  Debugging Information (artifact)}.
\newblock
\newblock
\urldef\tempurl%
\url{https://doi.org/10.5281/zenodo.10568392}
\showDOI{\tempurl}


\bibitem[Tice and Graham(1998)]%
        {ticeOPTVIEWNewApproach1998}
\bibfield{author}{\bibinfo{person}{Caroline Tice} {and}
  \bibinfo{person}{Susan~L. Graham}.} \bibinfo{year}{1998}\natexlab{}.
\newblock \showarticletitle{{{OPTVIEW}}: A New Approach for Examining Optimized
  Code}. In \bibinfo{booktitle}{\emph{Proc. of PASTE '98}}.
\newblock
\urldef\tempurl%
\url{https://doi.org/10.1145/277631.277636}
\showDOI{\tempurl}


\bibitem[Yang et~al\mbox{.}(2011)]%
        {yangFindingUnderstandingBugs2011}
\bibfield{author}{\bibinfo{person}{Xuejun Yang}, \bibinfo{person}{Yang Chen},
  \bibinfo{person}{Eric Eide}, {and} \bibinfo{person}{John Regehr}.}
  \bibinfo{year}{2011}\natexlab{}.
\newblock \showarticletitle{Finding and Understanding Bugs in {{C}} Compilers}.
  In \bibinfo{booktitle}{\emph{Proc. of PLDI '11}}.
\newblock
\urldef\tempurl%
\url{https://doi.org/10.1145/1993498.1993532}
\showDOI{\tempurl}


\bibitem[Zellweger(1983)]%
        {zellwegerInteractiveHighlevelDebugger1983a}
\bibfield{author}{\bibinfo{person}{Polle~T. Zellweger}.}
  \bibinfo{year}{1983}\natexlab{}.
\newblock \showarticletitle{An Interactive High-Level Debugger for Control-Flow
  Optimized Programs}. In \bibinfo{booktitle}{\emph{Proc. of SIGSOFT '83}}.
\newblock
\urldef\tempurl%
\url{https://doi.org/10.1145/1006147.1006183}
\showDOI{\tempurl}


\bibitem[Zellweger(1984)]%
        {zellwegerInteractiveSourcelevelDebugging1984}
\bibfield{author}{\bibinfo{person}{Polle~T. Zellweger}.}
  \bibinfo{year}{1984}\natexlab{}.
\newblock \emph{\bibinfo{title}{Interactive Source-Level Debugging of Optimized
  Programs}}.
\newblock \bibinfo{thesistype}{Ph.\,D. Dissertation}.
  \bibinfo{school}{University of California, Berkeley}.
\newblock
\urldef\tempurl%
\url{https://search.library.berkeley.edu/permalink/01UCS_BER/1thfj9n/alma991002570669706532}
\showURL{%
\tempurl}


\bibitem[Zurawski and Johnson(1991)]%
        {zurawskiDebuggingOptimizedCode1991}
\bibfield{author}{\bibinfo{person}{Lawrence~W Zurawski} {and}
  \bibinfo{person}{Ralph~E Johnson}.} \bibinfo{year}{1991}\natexlab{}.
\newblock \bibinfo{title}{Debugging Optimized Code with Expected Behavior}.
\newblock
\newblock
\newblock
\shownote{Unpublished draft}.


\end{thebibliography}

\end{document}